\documentclass[aps,prb,longbibliography,showpacs,twocolumn,superscriptaddress,amsmath,amssymb]{revtex4-2}
 
 \usepackage[utf8]{inputenc}
 \usepackage{amsmath,amsfonts,amssymb}
 \usepackage{pifont}
 \usepackage{rotating}
 \usepackage{graphicx}    
 \usepackage{epstopdf}
 \usepackage{color}
 \usepackage[caption=false]{subfig}
 \usepackage{soul}
 \usepackage{booktabs}
 \usepackage{hyperref}
 \definecolor{dark-red}{rgb}{0.9,0.15,0.15}
 \definecolor{dark-blue}{rgb}{0.15,0.15,0.4}
 \definecolor{medium-blue}{rgb}{0,0,0.5}
 \hypersetup{colorlinks=true,citecolor=blue,linkcolor=blue,urlcolor=blue}

\begin{document}
\title{Observation of a gapped phase in the one-dimensional $S = {\frac{1}{2}}$ Heisenberg antiferromagnetic chain Cu(Ampy)ClBr}
\author{Saikat Nandi}
\email{saikatnandi9@gmail.com}
\affiliation{Department of Physics, Indian Institute of Technology Bombay, Mumbai 400076, India}
\author{Monika Jawale}
\affiliation{Department of Physics, Indian Institute of Technology Bombay, Mumbai 400076, India}
\author{Sanjay Bachhar}
\affiliation{Department of Physics, Indian Institute of Technology Bombay, Mumbai 400076, India}
\affiliation{Department of Physics, Duke University, Durham, North Carolina 27708, USA}
\author{Rahul Kumar}
\affiliation{School of Advanced Materials, and Chemistry and Physics of Materials Unit, Jawaharlal Nehru Centre for Advanced Scientific Research, Bangalore-560064, India}
\author{Marlis Schuller}
\affiliation{Experimental Physics V, Center for Electronic Correlations and Magnetism, University of Augsburg, D-86135 Augsburg, Germany}
\author{Rabindranath Bag}
\affiliation{Department of Physics, Duke University, Durham, North Carolina 27708, USA}
\author{J. M. Wilkinson}
\affiliation{ISIS Pulsed Neutron and Muon Source, STFC Rutherford Appleton Lab, Harwell Campus, Didcot OX11 0QX, Oxon, England}
\author{Jörg Sichelschmidt}
\affiliation{Max Planck Institute for Chemical Physics of Solids, 01187 Dresden,
Germany}
\author{A. Sundaresan}
\affiliation{School of Advanced Materials, and Chemistry and Physics of Materials Unit, Jawaharlal Nehru Centre for Advanced Scientific Research, Bangalore-560064, India}
\author{Sara Haravifard}
\affiliation{Department of Physics, Duke University, Durham, North Carolina 27708, USA}
\author{N. Büttgen}
\affiliation{Experimental Physics V, Center for Electronic Correlations and Magnetism, University of Augsburg, D-86135 Augsburg, Germany}
\author{A.V. Mahajan} 
\email{mahajan@phy.iitb.ac.in}
\affiliation{Department of Physics, Indian Institute of Technology Bombay, Mumbai 400076, India}

 \begin{abstract}
 
  Spin-1/2 Heisenberg antiferromagnetic frustrated spin chain systems display exotic ground states with unconventional excitations and distinct quantum phase transitions as the ratio of next-nearest-neighbor to nearest-neighbor coupling is tuned. 
  We present a comprehensive investigation of the structural, magnetic, and thermodynamics properties of the spin-1/2  compound, Cu(Ampy)ClBr (Ampy= C$_6$H$_8$N$_2$ = 2-(Aminomethyl)pyridine) via x-ray diffraction, magnetization, specific heat, $^1$H nuclear magnetic resonance (NMR), electron spin resonance (ESR), and muon spin relaxation ($\mu$SR) techniques. The crystal structure features an anisotropic triangular chain lattice of magnetic Cu$^{2+}$ ions. Our bulk and local probe experiments detect neither long-range magnetic ordering nor spin freezing down to 0.06 K despite the presence of moderate antiferromagnetic interaction between Cu$^{2+}$ spins as reflected by a Curie-Weiss temperature of about $-9$ K from the bulk susceptibility data. A broad maximum is observed at about 9 K in magnetic susceptibility and specific heat data, indicating the onset of short-range spin correlations. At low temperatures, the zero-field magnetic specific heat and the $^1$H NMR spin-lattice relaxation rate follow an exponential temperature dependence, indicating the presence of gapped magnetic excitations.
  Furthermore, persistent spin dynamics down to 0.088 K observed by zero-field $\mu$SR evidences lack of any static magnetism.

\end{abstract}
				
\maketitle

\section{Introduction}
 
   Low-dimensional frustrated spin systems, where a macroscopic number of quasi-degenerate states compete, provide a novel route for exploring emergent exotic phenomena at low temperatures \cite{vasiliev2018milestones, zapf2014bose}. Moreover, due to the reduced dimensionality, an interplay of exchange coupling and strong quantum fluctuations could trigger disorder induced spontaneous symmetry  breaking and give rise to unusual physical phenomena \cite{giamarchi2003quantum}. Both the magnetic N$\acute{e}$el state, and the valence bond solids (VBS) are states of two different broken symmetries, the former breaks spin-rotation symmetry, and the latter breaks spatial symmetry of lattice. When quantum fluctuations between a macroscopic number of quasi-degenerate states hinder the selection of particular order, one might end up with various kinds of spin liquids. The $S = 1/2$ Heisenberg antiferromagnetic (HAF) chain is the simplest example, and its ground state is characterized as a Tomonaga-Luttinger liquid (TLL), where spin-spin correlations are expected to show power-law decays with distance and there is the presence of fractionalized excitations \cite{tomonaga1950remarks,luttinger1963exactly, halden1980}. 
    
   A simple realization of structured one-dimensional (1D) frustrated system is the Heisenberg spin-$1/2$ $J_1-J_2$ chain system consisting of nearest-neighbor (NN) $J_1$ and next-nearest-neighbor (NNN) $J_2$ couplings. This model has been studied in detail to obtain the magnetic phase diagram. For the antiferromagnetic $J_1-J_2$ model for a spin-1/2 chain, the ground state is a singlet  for any $\alpha=J_2/J_1$. The $\alpha=0$ limit is the gapless HAF with a nondegenerate ground state (TLL) while the degenerate ground state at $\alpha=1/2$ corresponds to the  Majumdar-Ghosh (MG) point \cite{majumdarghosh1969}.   
  The TLL-dimer transition occurs at $\alpha=$ 0.2411 \cite{okamoto1992fluid}, the expected spin gap is exponentially small for $\alpha=$ 0.3. The dimer-gapped phase starts to develop rapidly for $\alpha \approx$ 0.4-0.5. A random-bond 1D spin-1/2 Heisenberg model with $J_1-J_2$ couplings  \cite{uematsu2021frustration} was investigated by Uematsu \textit{et al}.  where the ground-state phase diagram was constructed in the randomness versus  frustration plane ($J_2/J_1$).
  Various spin chain systems with  $J_1-J_2$ interactions have been investigated in the past. A few examples of $S=1/2$ Cu-based zigzag chain systems (3$d^9$) include the investigation of the 
  spin-Peierls transition in
  CuGeO$_3$ \cite{hase1993,riera1995,hirota1994,pouget1994}, incommensurate helimagnetism in LiCu$_2$O$_2$, NaCu$_2$O$_2$ \cite{masuda2004, capogna2005}, and observation of the MG point in Cu$_3$(MoO$_4$)(OH)$_4$ \cite{stefan2017}.
  On the other hand, Sr$_2$CuO$_3$ is a nearly ideal realization of the spin-1/2 uniform HAF chain. It has a large exchange energy ($\sim$ 2200 K) compared to it's 3D long range ordering (LRO) temperature ($\sim$ 5 K) \cite{ami1995, motoyama1996,takigawa1997}.
  Metal-organic frameworks (MOFs) have some advantages compared to the usual inorganic systems for such studies because the exchange couplings, spin gap, and the ground state properties can often be tuned by engineering the synthesis route and using appropriate organic ligands.
           
    Herein, we focus on Cu(Ampy)ClBr (Ampy =  2-(Aminomethyl)pyridine = C$_6$H$_8$N$_2$), which belongs to the family of Cu(Ampy)X$_2$ (X= Cl, Br) with zigzag Cu-Cu magnetic chains \cite{OConnor1985, helis1977,kikuchi2000}. No detailed structural, magnetic, and thermodynamic studies have yet been reported in Cu(Ampy)ClBr.  
   Cu(Ampy)Br$_2$ shows LRO around $T \approx 4$ K indicated from a bifurcation in the zero field cooled and field cooled susceptibility data (see \cite{Cu(Ampy)ClBr_Supp} for details), whereas in the case of  Cu(Ampy)Cl$_2$ \cite{OConnor1985} the magnetic susceptibility data deviate from the 1D chain model below 2.5 K and
   	show a small kink around 2.42 K. The specific heat data (see \cite{Cu(Ampy)ClBr_Supp} for details) on  Cu(Ampy)Cl$_2$ have a sharp anomaly around 0.7 K. We aim to generate randomness in the nearest neighbor exchange by a 50:50 halide (Cl/Br) mixing in this system. In addition, a non-zero $J_2$ is likely to be present in this case. In case of Cu(Ampy)Br$_2$, a value of $J_2/ J_1 = 0.2$ was obtained using the exact diagonalization method, and no spin gap was observed \cite{kikuchi2000}. The title compound Cu(Ampy)ClBr has a structure similar to that of Cu(N$_2$H$_5$)Cl$_3$, where the ratio of the couplings is estimated to be $J_2/ J_1 \sim 0.25$ \cite{hagiwara2001}. 
           
      In this paper, we present a  detailed investigation of the $S=1/2$ anisotropic zigzag magnetic chain system Cu(Ampy)ClBr through powder X-ray diffraction (XRD), magnetic susceptibility $\chi (T)$, specific heat, nuclear magnetic resonance (NMR), electron spin resonance (ESR), and muon spin relaxation ($\mu$SR)  measurements. The interaction between the Cu$^{2+}$ moments is predominantly antiferromagnetic in nature, as evidenced by the Curie-Weiss (CW) temperature $\theta_{CW} \approx -9$ K. Our experimental results reveal no LRO down to 0.06 K and instead show a broad maximum around $T\sim 9$ K, suggesting the presence of short-range AFM correlations. The magnetic specific heat $C_{mag}$ data and $^1$H NMR spin-lattice relaxation rate ($1/T_1$) show  gapped excitations at low temperatures. Our $\mu$SR analysis reveals the presence of two components in the muon depolarization, arising from the coexistence of rapidly and slowly fluctuating Cu$^{2+}$ moments. The faster component (which we suggest comes from muon sites near the chain-ends) follows the Redfield variation with field whereas the other (which we suggest is from muon sites in the middle of the chain) appears to vary as for diffusive motion of spinons. 
      
      \begin{figure}[ht]
      	\begin{center}
      		\includegraphics[width=0.95\columnwidth]{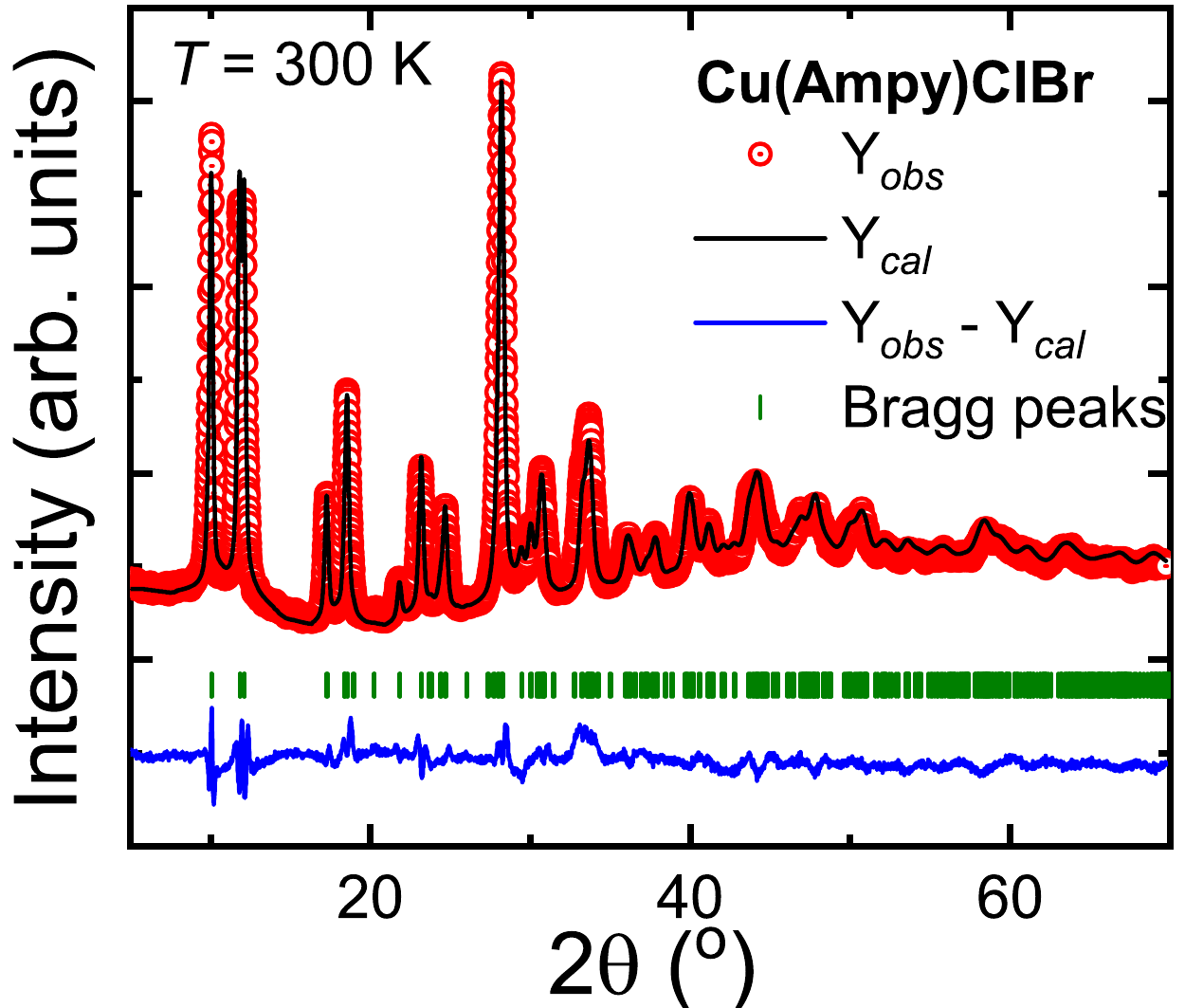}
      		\caption{Rietveld refinement of the powder x-ray diffraction pattern for Cu(Ampy)ClBr at 300 K. The red circles represent the observed (Y$_{obs}$) intensity, whereas the calculated (Y$_{cal}$) patterns and the difference (Y$_{obs}-$Y$_{cal}$) are shown in black and blue lines, respectively. The green tick marks denote the allowed Bragg positions.}
      		\label{F1}
      	\end{center}
      \end{figure}
  \section {Experimental Details}
 Polycrystalline samples of Cu(Ampy)ClBr were synthesized in the following manner. We added 2 equivalent of 2-(Aminomethyl)pyridine (Ampy) in 4 ml of methanol to 1 equivalent of CuCl$_2$.2H$_2$O in 2 ml of methanol and 1 equivalent of anhydrous CuBr$_2$ in 2 ml of methanol. After 30 minutes of stirring at room temperature, a significant quantity of green precipitate formed. The precipitate was separated by filtration followed by cleaning with methanol and then dried for 24 h. The phase purity of the sample was checked by doing powder x-ray diffraction (XRD) measurements at room temperature with a high resolution Rigaku diffractometer using Cu-K$_\alpha$ radiation ($\lambda$ = 1.5406 Å).   
 
 The magnetic susceptibility was measured as a function of temperature ($0.4 \leq T\leq 300 $ K), in  $H$ = 100 Oe and 10 kOe. The isothermal magnetization ($M$ \textit{vs} $H$) was also measured at different temperatures. Further, a SQUID Vibrating Sample Magnetometer cooled with liquid $^3$He (iQuantumHe3, Quantum Design) was used for measurement down to 0.4 K.  Specific heat measurements were performed as a function of temperature using a standard relaxation technique with a physical property measurement system (PPMS, Quantum Design), on a sintered pellet (8.2 mg) under magnetic fields $0\leq H \leq 90 $ kOe. The contribution of the addenda was measured in separate runs and subtracted. In addition, low-temperature specific heat measurements were carried out down to 0.06 K for several applied magnetic fields $0\leq H \leq 120 $ kOe, using a dilution refrigerator insert in a PPMS Dynacool (Quantum Design). Zero field (ZF) and longitudinal field (LF) (i.e., a field directed parallel to the initial muon-spin direction) muon spin relaxation ($\mu$SR) measurements were carried out using the MUSR spectrometer at the ISIS Neutron and Muon Source (STFC Rutherford Appleton Laboratory, UK). The spectra were collected  from 0.088 K  to 4 K using a dilution refrigerator.  A powder sample (thickness $\sim$ 2 mm) was mounted directly on a silver backing plate (99.995+$\%$ purity) using GE varnish and was covered with a thin silver foil. The $^{1}$H nuclear magnetic resonance (NMR) measurements have been performed in a swept field magnet (Cryomagnetics Inc.) with a Tecmag spectrometer down to 1.7 K. NMR measurements down to 0.03 K were also performed on $^{1}$H using a dilution refrigerator. From our measurements, we obtained $^{1}$H NMR spectra (down to 0.03 K) and the $^1$H spin-lattice relaxation rate  ($1/T_1$) (down to 0.09 K) as a function of temperature. Also, electron spin resonance (ESR) measurements were carried out for $3\leq T\leq 300$ K using a helium-flow cryostat and a continuous wave ESR spectrometer at X-band frequencies ($\nu = 9.4$ GHz).
\section {Experimental Results}
\subsection{Crystal structure}
\begin{figure}[ht]
	\begin{center}
		\includegraphics[width=0.95\columnwidth]{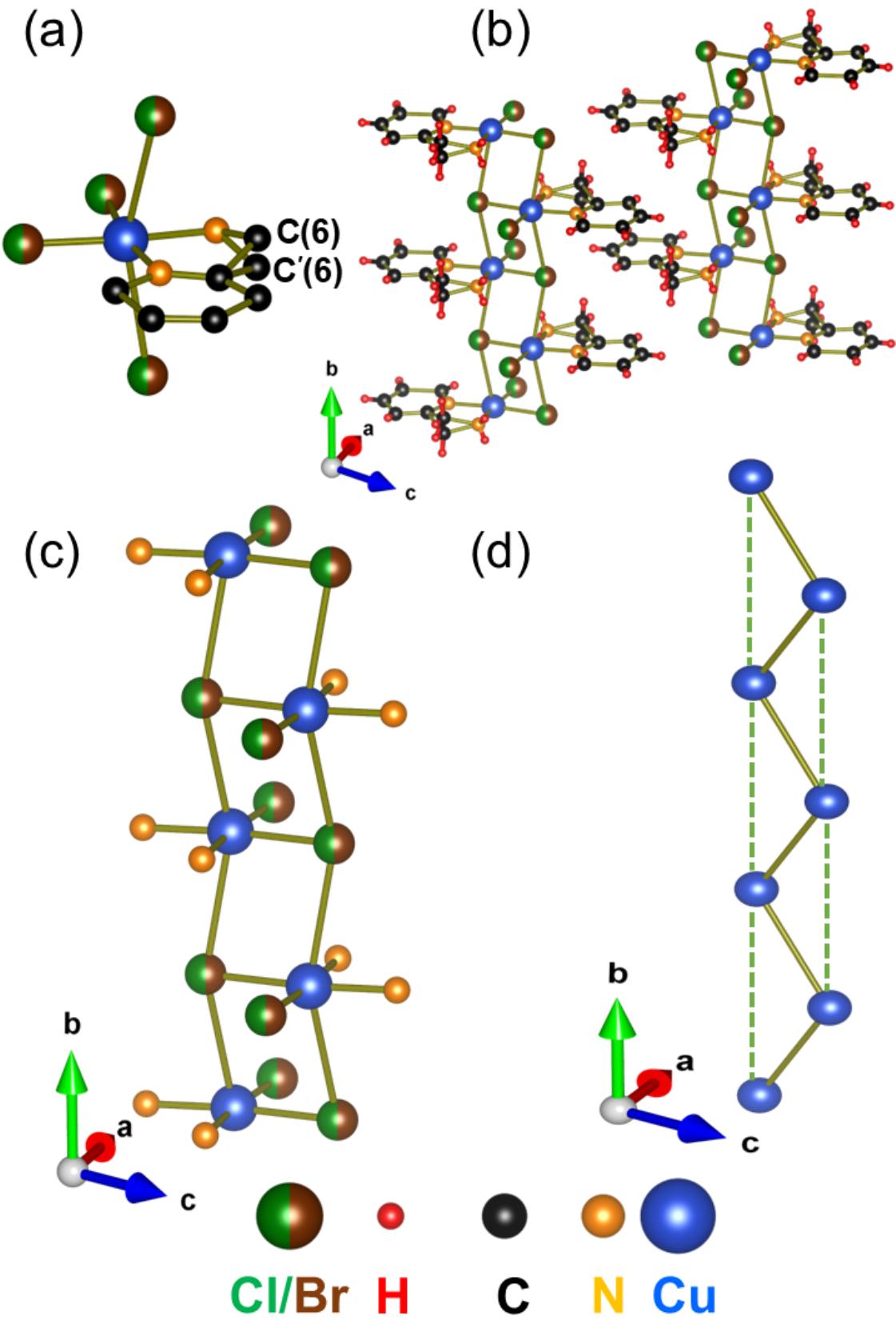}
		\caption{(a) Schematic diagram of Cu(Ampy)ClBr molecular unit with two positions of C(6) atom. Hydrogen atoms are omitted for clarity. Each unit is bridged to form an infinite chain-like structure. (b) Molecular structure of Cu(Ampy)ClBr. (c) View of the bridging network in Cu(Ampy)ClBr running along the crystallographic $b$ axis. (d) The possible exchange couplings between Cu ions forming zigzag spin-chain structure. The solid and dashed lines denote the nearest neighbor coupling and the next-nearest neighbor coupling respectively.}
		\label{F2}
	\end{center}
\end{figure}

Powder XRD data collected at room temperature for Cu(Ampy)ClBr are shown in Fig. \ref{F1}. Rietveld refinement of the XRD patterns was carried out using FULLPROF software package~\cite{rodriguez1993recent}, taking initial structure parameters from Ref. \cite{OConnor1985, helis1977}. The result suggests that the obtained structure of Cu(Ampy)ClBr is similar to its analogous coordination compound Cu(Ampy)X$_2$ (X=Cl, Br), reported previously \cite{OConnor1985, helis1977}. During refinement, the positions of carbon, hydrogen, nitrogen atoms for Cu(Ampy)ClBr could not be refined and were kept fixed to the values of  Cu(Ampy)Cl$_2$. The lattice constants $a$, $b$, $c$ and $\beta$ for Cu(Ampy)ClBr are found to be between those of the pure Cl and Br analogs. The refined structural parameters are tabulated in Table \ref{tab:table1}.
 Carbon atom C(6) of the Ampy rings are disordered over two equally-populated positions separated by 0.534(1) \AA~ across the mirror plane. In Fig. \ref{F2}(a), these two positions are denoted as C(6) and C$^\prime$(6). Such disorder was also observed in other Cu$^{2+}$ complexes Cu(Ampy)Cl$_2$ \cite{OConnor1985}, Cu$(en)$Cl$_2$ ($en =$ C$_2$H$_{8}$N$_2$ = ethylenediamine) \cite{harvey1986}, Cu$(tn)$Cl$_2$ ($tn =$ C$_3$H$_{10}$N$_2$ = 1,3-diaminopropane) \cite{tarasenko2023}. 

  Fig. \ref{F2}(b) presents the Cu(Ampy)ClBr molecular structure of Cu(Ampy)ClBr with Ampy= C$_6$H$_8$N$_2$ ligand. 
Shown in Fig. \ref{F2} (c) is the coordination around each Cu$^{2+}$ ion is a tetragonally distorted octahedron due to the Jahn-Teller (J-T) effect. Due to the J-T distortion, magnetic interaction between spin-1/2 Cu$^{2+}$ moments are expected to be anisotropic. The four in-plane short bonds are formed by two cis-nitrogen (N) atoms of the C$_6$H$_8$N$_2$ ligand and two halide (X $=$ Cl$/$Br) atoms, X(1) and X(2). The out-of-plane bonds are completed by two X(2) atoms at much higher distances from Cu$^{2+}$ ion. In the basal plane, the Cu-X bond distance varies from 2.374(1) to 2.511(1) Å while the longest apical bond Cu-X(2) distance is 3.249(0) Å. The Cu-N bond lengths are in the range of 1.940-2.037 Å. The nearest Cu-Cu distance is 3.732(1) \AA. The zigzag chains are formed by incorporating the second NN, with a Cu-Cu distance = 6.324(1) \AA (see Fig. \ref{F2}(d)). The zigzag Cu-Cu magnetic chains propagate along the $b$-axis with the  non-linear Cu-X(2)-Cu-X(2)-Cu superexchange pathway of magnetic interactions (see Fig. \ref{F2}(c)). Non-linear pathways propagate a weaker superexchange according to the Goodenough-Kanamori rules \cite{goodenough1963}. Superexchange through the Ampy, C$_6$H$_8$N$_2$ ring pathway orthogonal to the zigzag chain direction arises from the weak coupling $\pi$ mechanism and these Ampy ligands link the chains together into a 2D extended network. Magnetically, these $S= 1/2$ chains appear to be well isolated. 
\begin{figure*}
	\begin{center}
		\includegraphics[width=2.05\columnwidth]{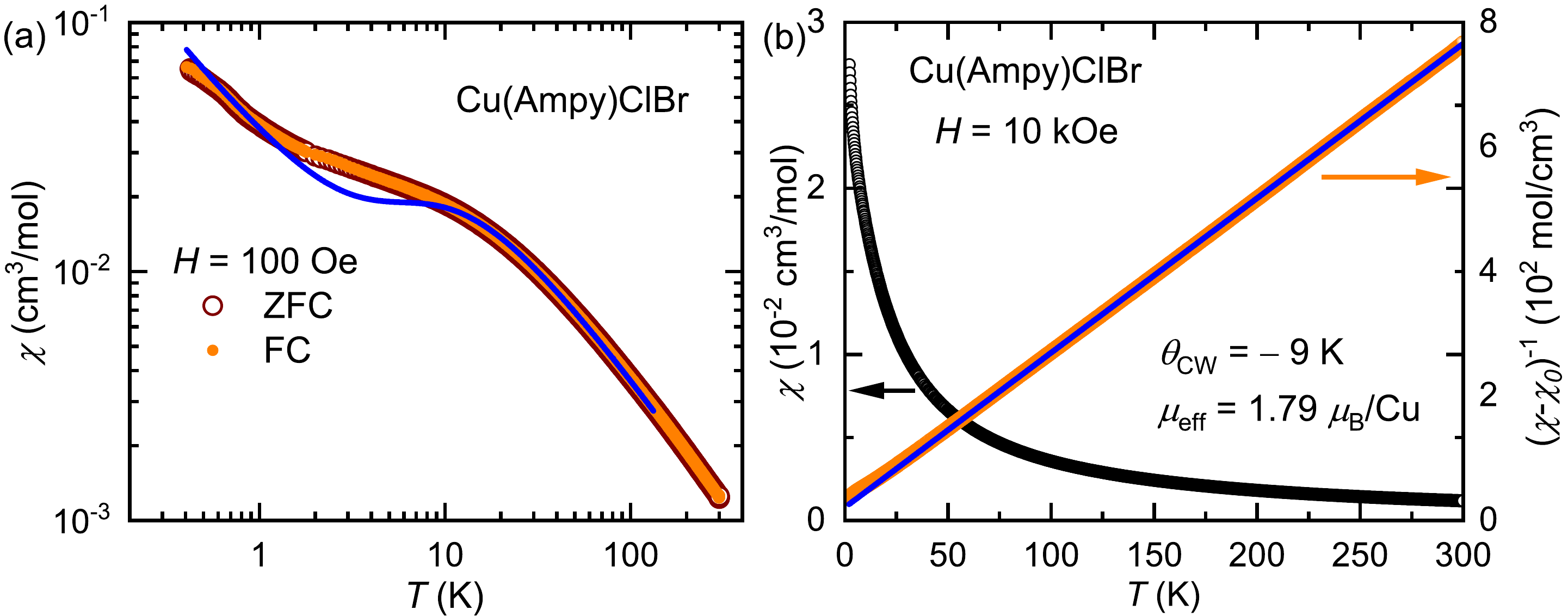}
		\caption{(a) Temperature dependence of the static magnetic susceptibility $\chi(T)$ measured at $H=$ 100 Oe in  field cooled (FC) (orange close symbols) and zero field cooled (ZFC) (brown open symbols) processes on a log-log scale. No bifurcation between ZFC and FC $\chi(T)$ in is seen down to 0.4 K. The solid blue line is the fit to Eq. (\ref{eq1}), as described in the text. (b) Left $y$-axis: Temperature dependence of the static magnetic susceptibility $\chi(T)$ measured at $H=$ 10 kOe. Right $y$-axis: The temperature dependence of inverse magnetic susceptibility data free from $\chi_{0}$. The Curie-Weiss fitting is shown by the solid blue line.}
		\label{F3}
	\end{center}
\end{figure*}

\begin{table}[h]
	\centering{}\caption{\label{tab:table1}{Lattice parameters and quality factors are obtained from the Rietveld refinement of the room-temperature powder XRD data of Cu(Ampy)ClBr (monoclinic, $P$2$_1/$$m$).}}
	\vspace{0.5cm}
	\begin{tabular}{lll}
		\hline
		\hline  
		Prameters  \tabularnewline
		\hline 
		Space group & $P$2$_1/$$m$ & \tabularnewline
		Lattice parameter  & $\mathit{a}$ = 8.147(1) \AA & \tabularnewline
		& $\mathit{b}$ = 6.324(0) \AA & \tabularnewline
		& $\mathit{c}$ = 9.569(1) \AA & \tabularnewline
		$\alpha$, $\beta$, $\gamma$ & $\alpha=\gamma=  $90$^{\mathrm{o}}, \beta=$113.379(8)$^{\mathrm{o}}$ & \tabularnewline
		Cell volume {[}\AA $^{3}${]} & 452.608(93) &  \tabularnewline
		R$_{\rm p}$, R$_{\rm wp}$, R$_{\rm exp}$(\%) & 21.1, 16.4, 14.6 &  \tabularnewline
		Bragg R-factor(\%) & 15.7 & \tabularnewline
		RF-factor(\%) & 16.1 &  \tabularnewline
		$\chi^{2}$ & 1.26 & \tabularnewline
		\hline
		\hline 
	\end{tabular}
\end{table}
\subsection{Magnetization}
The temperature dependence of the magnetic susceptibility $\chi (T)$ ($= M/H$) of Cu(Ampy)ClBr measured in an applied field of $H =$ 100 Oe and 10 kOe is shown in Fig. \ref{F3}(a) and left $y$-axis of \ref{F3}(b) respectively. We observed no thermal hysteresis between the zero-field-cooled (ZFC) and field-cooled (FC) curves which indicates the absence of any frozen spins. No obvious signature of magnetic LRO is seen down to 0.4 K. At high temperatures, $\chi (T)$ follows the standard paramagnetic behavior and then, with decreasing temperature, it shows a broad and weak maximum around $T \simeq$ 9 K, qualitatively similar to other chain systems, characteristic of low-dimensional systems featuring short-range ordering.

We first analyze the high$-T$ susceptibility data ($150-300$ K) using the Curie-Weiss (CW) law $\chi (T) = \chi_0 + C/(T-\theta_{CW})$ (see Fig. \ref{F3}(b)). The value of temperature-independent susceptibility is $\chi_0 = -1.28 \times10^{-4}$ cm$^3$/mol, which includes van Vleck paramagnetic susceptibility $\chi_{VV}$ and core diamagnetic susceptibility $\chi_{dia}$. Adding the core diamagnetic susceptibilities of the individual ions \cite{Bain}, the calculated total  diamagnetic susceptibility $\chi_{dia}$ turns out to be $\thickapprox -1.4 \times10^{-4}$ cm$^3$/mol for Cu(Ampy)ClBr. Deducting  $\chi_{dia}$ from $\chi_{0}$ gives us the  Van Vleck susceptibility $\chi_{VV} \simeq 0.12 \times 10^{-4}$ cm$^3$/mol, resulting from the second derivative of the free energy under the influence of magnetic field. This is about ten times smaller than inferred values in other Cu-based systems. The CW fit yields $C =$ 0.40 cm$^3$ K/mol and a negative Curie-Weiss temperature of $\theta_{CW} \approx -9(1)$ K, which can be attributed to the predominance of antiferromagnetic (AF) exchange interactions between the Cu$^{2+}$ spins. The value of $C$ allows estimating the effective moment $\mu_{eff} = (3k_{B}C/N_{A}\mu ^{2}_{B})^{1/2}$ as 1.79 $\mu _B$/Cu, where $N_A$ is the Avogadro's number and $\mu_B$ is the Bohr magneton. This value of $\mu_{eff}$ is in good agreement with the theoretical value (1.73 $\mu _B$/Cu) for the free ion moment of $S = 1/2$ Cu$^{2+}$ ion.

In a mean field approach, using the experimental value of $\theta_{CW}$, one can estimate the average value of the AFM exchange coupling  as $\theta_{CW} = [-zJS(S+1)]/3k_B$ \cite{domb1964,guchhait2024}. Here, $J$ is the nearest-neighbour (NN) exchange coupling with the Heisenberg Hamiltonian $H = \sum J S_{i}\cdot S_{j}$ and $z$ denotes the number of NN spins associated with each Cu$^{2+}$
ion (see Fig. \ref{F1}(d)). In Cu(Ampy)ClBr, each Cu$^{2+}$ spin interacts with 2 neighboring Cu$^{2+}$ spins. Thus, using the value of $\theta_{CW}$, $z=2$, and $S=1/2$ in the above expression, we estimated $J/k_{B}=18(2)$ K. 

In order to differentiate between the spin chain $\chi_{chain} (T)$ and the impurity $\chi_{imp} (T)$ contributions, one can model $\chi (T)$ with the following expression,
 \begin{equation} \label{eq1}
	\chi (T) = \chi_{0} (T) + \chi_{imp} (T) + \chi_{chain} (T),
	\end{equation}
where $\chi_{0} (T)$ is the temperature independent susceptibility. Assuming that the low-$T$ rise in $\chi (T)$ data originates from the non-interacting paramagnetic impurity contributions, $\chi_{imp} (T) = C_{imp}/T$.
$\chi_{chain} (T)$  is the $S = 1/2$ uniform antiferromagnetic chain susceptibility expression provided by Johnston $et$ $al.$ \cite{johnston2000thermodynamics}.  Fitting of $\chi (T)$ data yields the intrachain interaction $J/k_B \simeq$ 14 K, $C_{imp} \simeq$ 0.027 cm$^3$ K/mol, and the Land$\acute{e}$ $g$-factor $g \simeq$ 2 (see Fig \ref{F3}(a)). The average value of the Cu-Cu exchange interaction $J/k_B$ compares well with the value estimated using $\theta_{CW}$ as expected for a $S=1/2$ spin-chain system. 

The contribution to the  low-$T$ tail in $\chi (T)$ includes one from presence of chain breaks \cite{granroth1998magnetic}.
\begin{figure}[ht]
	\begin{center}
		\includegraphics[width=0.95\columnwidth]{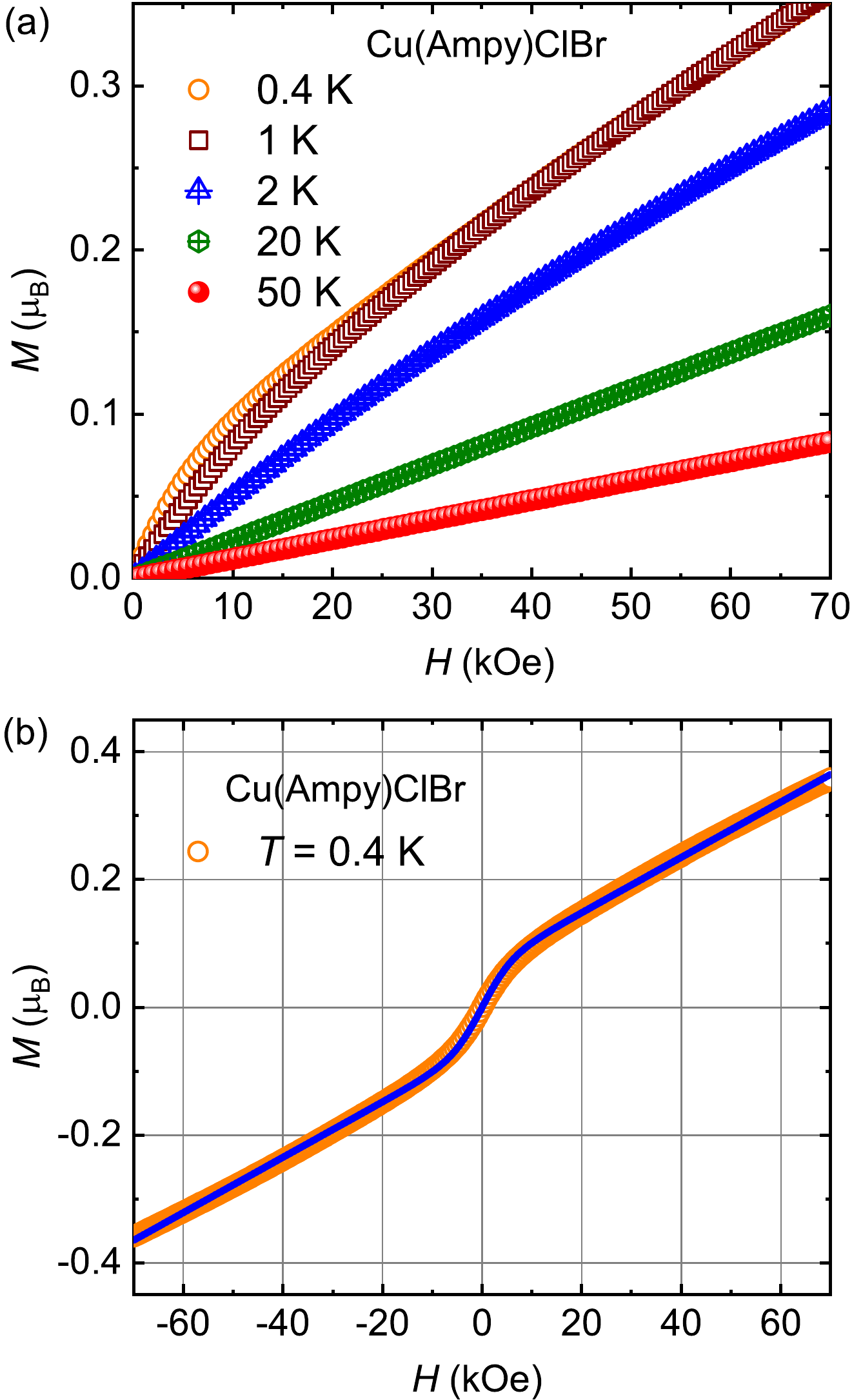}
		\caption{(a) Field dependent isothermal magnetization ($M$ \textit{vs} $H$) curve for the first quadrant at various temperatures. (b) Isothermal magnetization($M$) \textit{vs} $H$ curve at 0.4 K in the field range $-70$ kOe to 70 kOe. The absence of hysteresis indicates that there is no history dependence of magnetization in Cu(Ampy)ClBr. The solid line represents the fit obtained by combining the Brillouin function with a linear term.}
		\label{F4}
	\end{center}
\end{figure}

To examine the high-field magnetism of the compound, we have carried out isothermal magnetization ($M$ \textit{vs} $H$) at different temperatures in the range $0.4-50$ K (see Fig. \ref{F4}(a)) by gradually varying the magnetic field from $-70$ kOe to 70 kOe. We have not observed any signature of hysteresis in the isothermal curves (see Fig. \ref{F4}(b)). At high temperatures the $M$ increases linearly with $H$, as expected for an AF system. On the other hand, for $T=0.4$ K, the behavior is found to be nonlinear. To understand the origin of nonlinear magnetic isotherm at $T=0.4$ K, we used the combination of the Brillouin function with spin $S = 1/2$ spins and a linear function, i.e. $M(H)$ = $fN_{A}gS\mu_BB_{S}(g\mu_BSH/k_{B}T)+\chi H$, where $f$, $N_A$, $g$, and $k_B$ represent the fraction of paramagnetic impurities, the Avogadro constant, the  Land$\acute{e}$ $g$-factor, and the Boltzmann constant, respectively. The Brillouin function $B_S$, describes the behavior of paramagnetic spins as a function of magnetic field at a particular temperature \cite{kittel2005}. Our $T=0.4$ K magnetic isotherm data (see Fig. \ref{F4}(b)) analysis reveals that the Brillouin term arised from $\sim$ 6\% of $S=1/2$ spins with a $g$-factor of $\simeq$ 2. The slope of the high-field nonsaturated linear behavior of the magnetization curve at $T=0.4$ K yields $\chi=2.4 \times10^{-2}$ cm$^3$/mol, which is somewhat smaller than the value obtained by taking $M/H$ as in Fig. \ref{F3}(b) (note that the van Vleck susceptibility would be nearly two orders of magnitude smaller and would normally be evident from the slope of $M$ \textit{vs} $H$ at much higher fields).  Note that the slope of $M$ \textit{vs} $H$ is nearly unchanged from 1 K to 0.4 K indicating that the susceptibility is unchanged in this temperature range.   We find similar behavior in a 1D polymeric chain [Ni(HF$_2$)(3-Clpy)$_4$]BF$_4$ (py = pyridine) \cite{manson2012ni}. 

\begin{figure}[ht]
 	\begin{center}
 		\includegraphics[width=0.95\columnwidth]{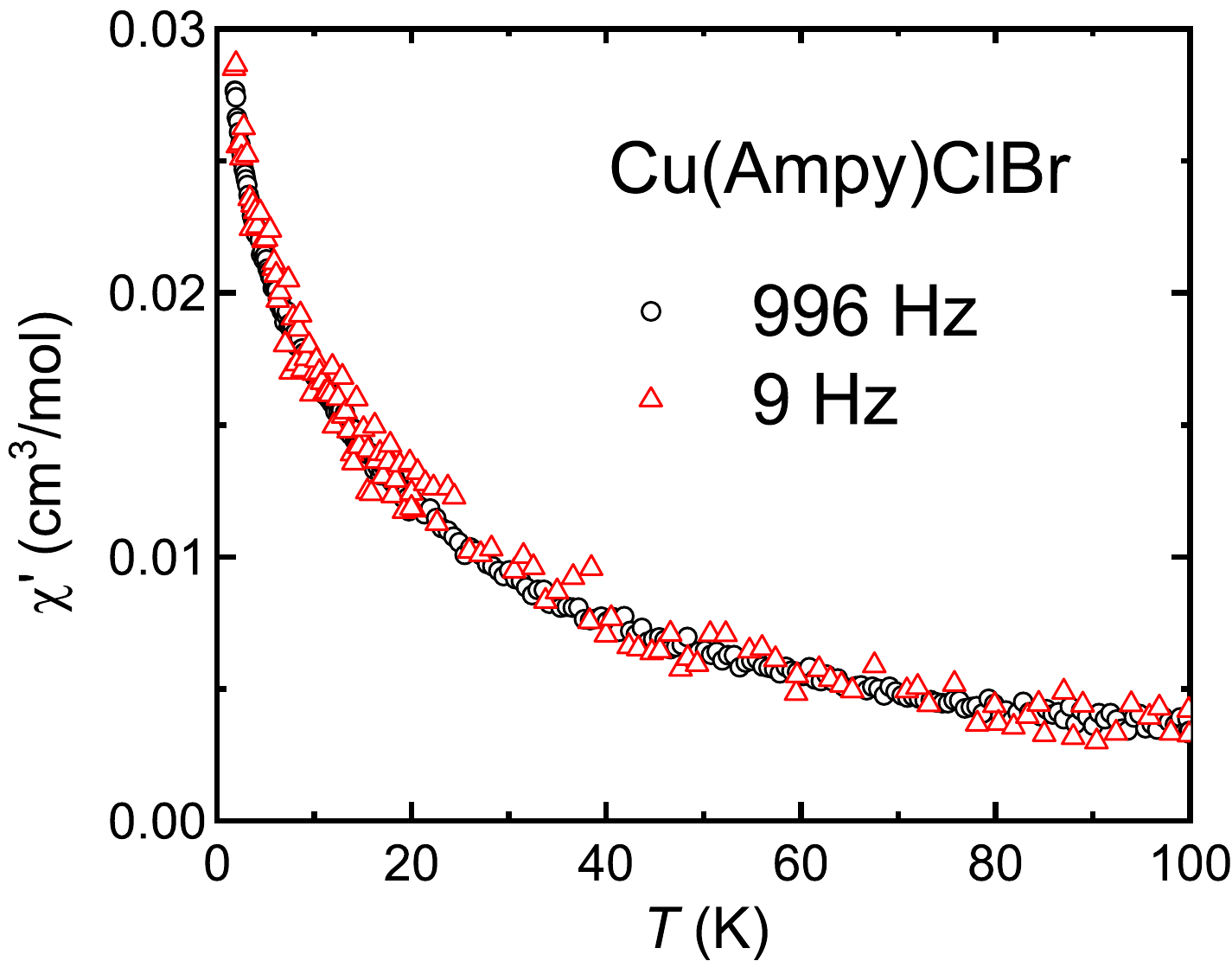}
 		\caption{Temperature dependence of the real component of the \textit{ac}-susceptibility $\chi^\prime$, measured at various frequencies. No frequency-dependent behavior can
 			be observed in the whole temperature range.}
 		\label{F5}
 	\end{center}
 \end{figure}

In addition, we have probed the spin dynamics down to 2 K through \textit{ac}-susceptibility measurements at  different frequencies in a small ac magnetic field ($H_{ac} = 1$ Oe). The inphase component of \textit{ac}-susceptibility ($\chi^{\prime}$) increases monotonically with decreasing temperature. Moreover, it shows a frequency-independent behavior, ruling out any spin freezing or spin-glass transition down to 2 K (see Fig. \ref{F5}). 
 \begin{figure*}
 	\begin{center}
 		\includegraphics[width=2.05\columnwidth]{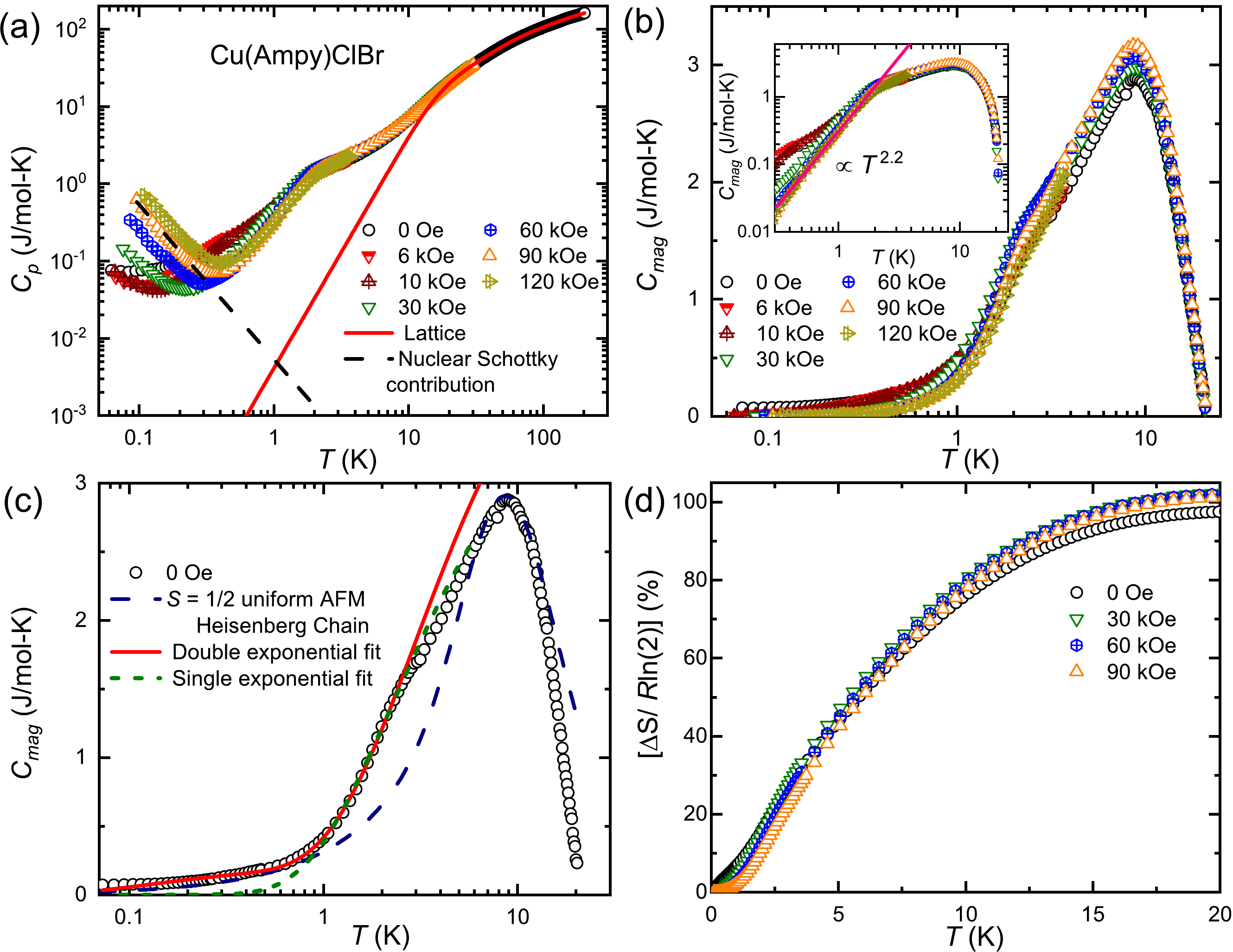}
 		\caption{(a) Temperature dependence of the specific heat ($C_{p}$) at different magnetic fields up to 120 kOe on a log-log scale. Red solid line ($C_{ph}$) is the fitted curve of the specific heat data measured in the absence of the magnetic field with the Debye-Einstein model. The black dashed line represents nuclear Schottky contribution ($C_{N}$) in an applied field of 90 kOe. (b) Magnetic contribution to the specific heat ($C_{mag}$) as a function of temperature in several magnetic fields. Inset: Power-law behavior ($C_{mag}\sim T^n$ with $n=2.2$) below about 2 K for fields above 30 kOe. (presented on a log-log scale for better visualization) (c)The temperature dependence of the zero-field magnetic specific heat, along with the curve predicted by the Johnston model for $J/k_{B} \sim 18$ K (dashed line). Also shown are the double-exponential fit (solid line) and the single-exponential fit (short dashed line) below $T=2.5$ K, as described in the text. (d) Magnetic entropy change as a function of temperature in various fields.}
 		\label{F6}
 	\end{center}
 \end{figure*}
\subsection{Specific Heat}
To further probe low-energy excitations of Cu(Ampy)ClBr, we performed specific heat measurements over the temperature range  $0.06 - 230$ K in different applied
fields. As shown in Fig. \ref{F6}(a), no sharp $\lambda$-type anomaly was observed down to the lowest temperature, indicating the absence of long-range magnetic order of Cu$^{2+}$ ions despite the presence of an AFM interaction ($\theta_{CW}=-9$ K). Due to the nuclear Schottky contribution, the low-temperature specific heat $C_p (T)$ data show upturns which shift to higher temperatures as the  magnetic field increases. At higher temperatures, the specific heat is dominated by phonon vibrations. In the absence of a nonmagnetic analog, we estimate the lattice specific heat $C_{ph} (T)$ data by fitting zero field $C_p (T)$ in the high-$T$ region using a combination of one Debye and three Einstein terms (see red solid line in Fig. \ref{F6}(a)) as given below,
\begin{equation}
	C_{\mathrm{Debye}}(T)=C_{\mathrm{D}}\left[9R\left(\frac{T}{\theta_{\mathrm{D}}}\right)^{3}\intop_{0}^{\theta_{\mathrm{D}/T}}\frac{x^{4}e^{x}}{(e^{x}-1)^{2}}dx\right]\label{eq:Debye}
	\end{equation}
	\begin{equation}
	C_{\mathrm{Einstein}}(T)=\sum C_{\mathrm{E_{j}}}\left[3R\left(\frac{\theta_{\mathrm{E}_{j}}}{T}\right)^{2}\frac{exp\left(\frac{\theta_{\mathrm{E}_{j}}}{T}\right)}{\left(exp\left(\frac{\mathrm{\theta}_{\mathrm{E}_{j}}}{T}\right)-1\right)^{2}}\right]\label{eq:Einstein},
\end{equation} where, $\theta_{\mathrm{D}}$, $\theta_{\mathrm{E}_{{j}}}$ ($j=$ 1, 2, and 3) are the Debye and Einstein temperatures, respectively and $\mathit{C_{\mathrm{D}}}$ and $\mathit{C_{{E_{j}}}}$ are the weightage factors corresponding to acoustic and optical modes of atomic vibrations, and $R$ is the universal gas constant. We obtained the weightage factors in the ratio $C_{\mathrm{D}}$:$C_{\mathrm{E_{1}}}$:$C_{\mathrm{E_{2}}}$:$C_{\mathrm{E_{3}}}$
$\equiv$ 2.15 : 3.00 : 3.45 : 10.10, resulting in $C_{\mathrm{D}} +\sum C_{ E_{j}} = 18.70$ which is very close to the expected 19 atoms per formula unit of Cu(Ampy)ClBr. 

The fitting also yields the Debye temperature $\theta_{\rm D}$= 100(2) K and the Einstein temperatures; $\theta_{\rm E_{1}}$=235(4) K, $\theta_{\rm E_{2}}$= 644(10) K and $\theta_{\rm E_{3}}$= 2068(55) K. 
This fit is then extrapolated down to the
lowest measured temperature and taken as the $C_{ph} (T)$.
The good fit at temperatures larger than about 20 K, as shown in Fig. \ref{F6}(a), suggests that these methods correctly obtain the phonon contribution to the specific heat in Cu(Ampy)ClBr. The experimental magnetic specific heat $C_{mag} (T)$ has been obtained by subtracting $C_{ph}(T)$ and the high temperature part of the nuclear Schottky contribution ($C_{N} \sim \alpha/T^2$) (see representative data in a 90 kOe field in Fig. \ref{F6}(a)) from the total specific heat.

Fig. \ref{F6}(b) depicts the magnetic specific heat $C_{mag}$ as a function of $T$ for different fields. $C_{mag}$ \textit{vs} $T$ features a broad hump (typical in low-dimensional systems) at around $\sim 9$ K. Such a hump is also suggested in the  $\chi(T)$ data.  
To fit the $C_{mag}$ data using the Johnston expression for the spin-1/2 HAF chain, we selected a range around the peak and then extrapolated the theoretical curve in both directions (see Fig. \ref{F6}(c)). The deduced value of $J/k_B$ ($\sim$ 18 K) is comparable to that obtained by fitting susceptibility data with the Johnston expression.  It is worth emphasizing that the fitted curve clearly departs from the experimental data below $T\sim 6$ K. Above $T\sim 13$ K, the $C_{mag} (T)$ data deviates from the model, a behavior which can be ascribed to the uncertainty in the estimation of $C_{ph} (T)$. 

Upon cooling, below $T\sim$ 2.5 K, $C_{mag} (T)$ in zero field show a rapid fall. To understand the nature of the excitations from the ground state, $C_{mag}$ data in zero field are fitted with the exponential function ($A_1 \exp(-\Delta_1 /k_{B}T)+A_2 \exp(-\Delta_2 /k_{B}T)$, where $A_1$ and $A_2$ are proportionality constants) over the temperature range 0.06 K $\leq T$ $\leq$ 2.5 K (see Fig. \ref{F6}(c)). The data were fitted to a double exponential, resulting in excitation gaps of $\Delta_1/k_{B}= 2.96(5) $ K and $\Delta_2/k_{B}=0.14(1) $ K, respectively. However, the fit below $T\sim$ 1 K with $\Delta_2/k_{B}=0.14(1)$ K is not quite satisfactory. We also fitted the data using a single exponential form, $\sim \exp(-\Delta /k_{B}T)$, with $\Delta/k_{B}=2.28(5)$ K; however, this deviates from activated behavior below 1 K (see Fig. \ref{F6}(c)). In any case, based on the $C_{mag}$ data analysis, it can be inferred that the Cu(Ampy)ClBr exhibits a gapped ground state. The magnetic specific heat below 2.5 K is suppressed on the application of a field and the variation changes from exponential to a power law. For fields above 30 kOe, the $C_{mag}$ data  below 2 K follow a power-law variation ($C_{mag}\sim T^n$) with $n=2.2$, as shown in Fig. \ref{F6}(b) (inset). This might be due to the closing of the gap with increasing field.

The respective magnetic entropy change $\Delta S$ in various fields are obtained by integrating $C_{mag}/T$ data from the base temperature and shown in Fig. \ref{F6}(d). At 20 K, the saturation value of $\Delta S$ is almost equal to the theoretical value $R$ln2, expected for $S=1/2$, irrespective of the value of applied field, where $R$ is the ideal-gas constant.

The main findings from our specific heat results are that we did not observe any magnetic LRO down to 0.06 K in Cu(Ampy)ClBr and we observe a ground state with gapped excitations. 
\begin{figure}[ht]
	\begin{center}
		\includegraphics[width=0.97\columnwidth]{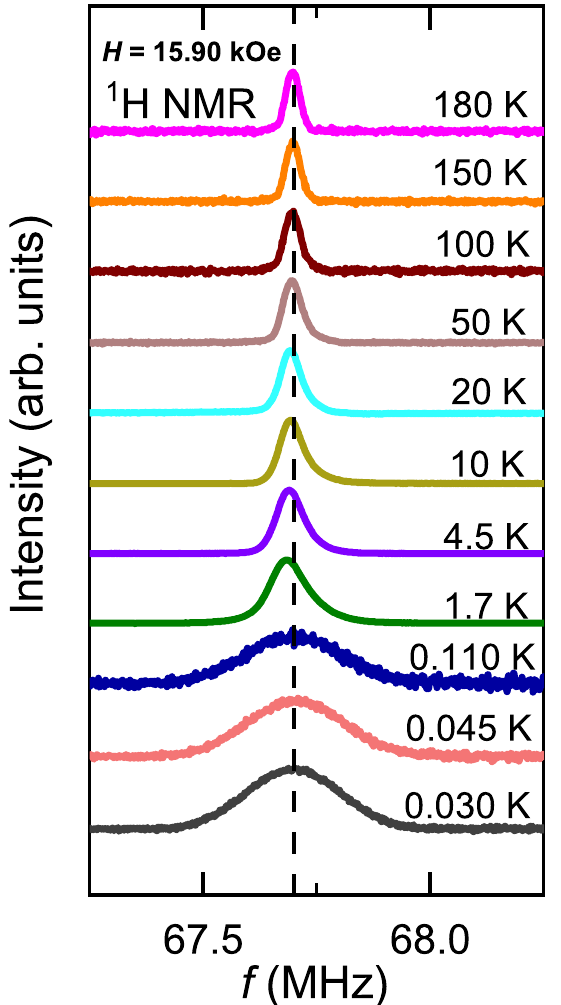}
		\caption {Temperature evolution of $^{1}$H NMR spectra at $H=15.90$ kOe in the temperature range $0.03-180$ K. The black dotted line is the reference position.}
		\label{F7}	
	\end{center}
\end{figure}
\subsection{Nuclear Magnetic Resonance}

Bulk magnetization at low temperatures  in many of the $S=1/2$ HAF chains is often  dominated by extrinsic paramagnetic impurities and/or chain-end effects. Therefore, to verify the intrinsic nature of Cu$^{2+}$ moments in Cu(Ampy)ClBr, $^1$H NMR($I=1/2$, $\gamma_N /2\pi = 42.57748$ MHz/T, and 99.98 \% natural abundance) was measured on the bulk sample. So by measuring the $^1$H NMR lineshape and spin-lattice relaxation rate $1/T_1$, one can extract static and dynamic properties associated with the magnetic spin chain formed by Cu$^{2+}$ ions. The $^1$H NMR signal was obtained using a spin-echo pulse sequence.
All our spectra were obtained from Fast Fourier transformation (FFT) technique by transforming half of the spin-echo measured in the time domain into the frequency domain. Fig. \ref{F7} shows the $^1$H-NMR spectra measured in the temperature range $0.030-180$ K for the frequency $f = 67.70$ MHz at an applied field of 15.90 kOe. In the present compound, there are 8 protons with two different Wyckoff positions, $2e$ ($x, 1/4, z$) and $4f$ ($x, y, z$) in a unit cell. 
 
We observed a single  spectral line with a FWHM of 31 kHz around zero shift position at 180 K.  The increase in FWHM with decreasing temperature can be attributed to the increase of bulk $\chi$ and consequently a greater distribution of local susceptibilities.
    The position of the central line remains unchanged with temperature, confirming a weak hyperfine coupling of $^1$H with the Cu$^{2+}$ ions due to a negligible overlap of their orbitals. The broadening of the spectral line is mainly due to the classical dipolar effects.
 
\begin{figure}[ht]
	\begin{center}
		\includegraphics[width=0.97\columnwidth]{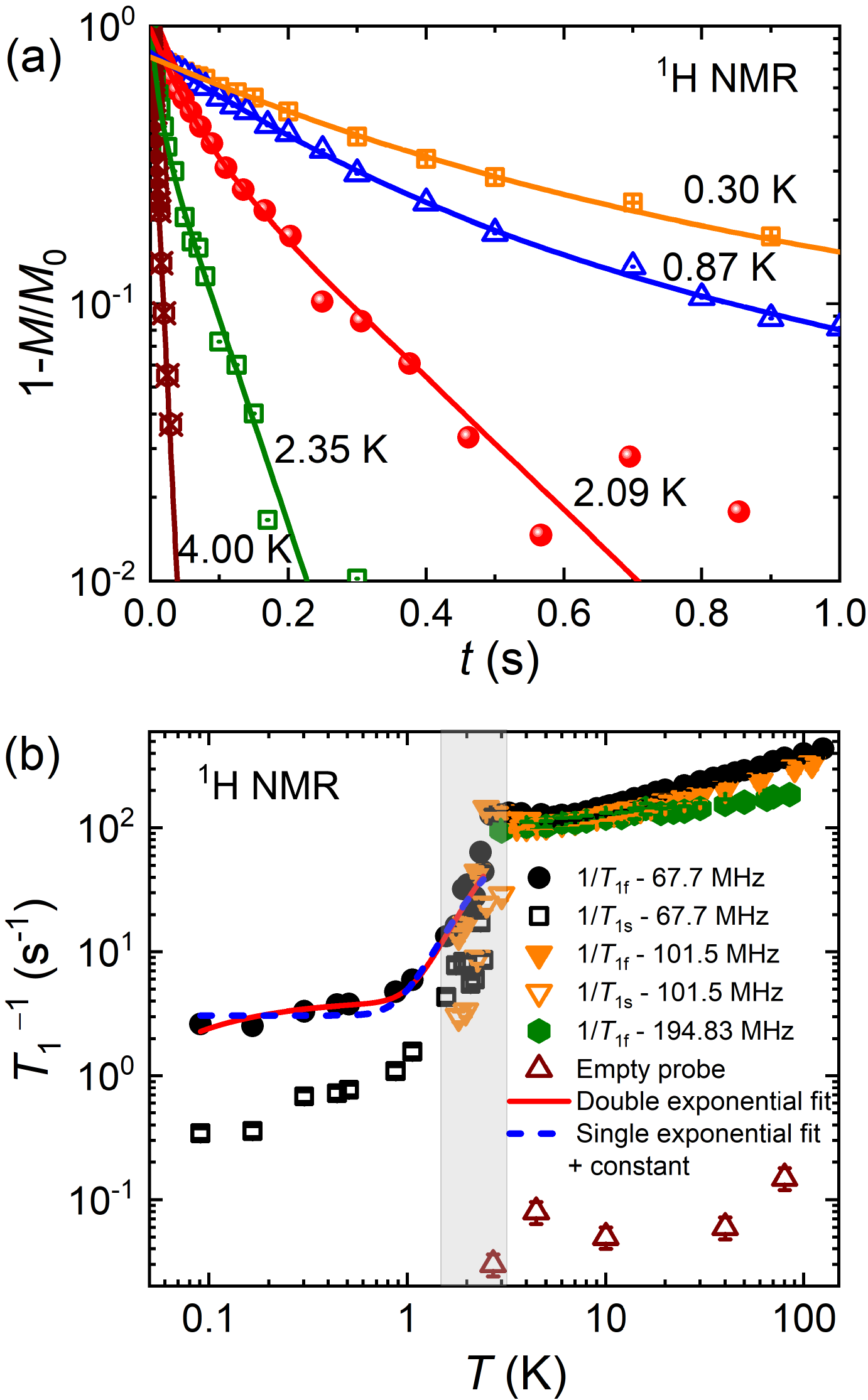}
		\caption {(a) The recovery of the longitudinal nuclear magnetization of $^{1}$H, measured at 67.7 MHz, as a function of time delays $t$ is presented in semilogarithmic scale for some selected temperatures and solid lines are fits using Eq. (\ref{eq4}) and (\ref{eq5}) as described in the text. (b) $^{1}$H spin-lattice relaxation rate ($1/T_1$) as a function of $T$ at 67.7 MHz, 101.5 MHz and 194.83 MHz (closed symbol: faster component, open symbol: slower component). The solid line represents the best-fit curve obtained using the double exponential behavior, and the short dashed line indicates a single-exponential fit with a constant term, as described in the text. The variation of $1/T_1$ for empty probe is shown by brown open triangle. For a clear visualization, the data are shown on a log-log scale.}
		\label{F8}	
	\end{center}
\end{figure}
To investigate the dynamical properties of Cu$^{2+}$ moments and the ground state properties, temperature-dependent $^1$H spin-lattice relaxation rates
($1/T_1$) were measured at different frequencies. We used the saturation recovery method for the $1/T_1$ measurements. For an $I=1/2$ nucleus, the recovery of the longitudinal nuclear magnetization $M(t)$ at a time $t$ is expected to follow a single exponential behavior after a saturating pulse sequence with the following form:
\begin{equation} \label{eq4}
	1-\frac{M(t)}{M(0)} = Ae^{-(t/T_{1})},
\end{equation}
where $M_0$ is the saturation magnetization as $t \to \infty$. In the high temperature regime (2.5 K $<T < 100$ K), $1/T_1$ at each $T$ was determined by fitting the data to Eq. (\ref{eq4}) at different frequencies (67.7 MHz, 101.5 MHz, and 194.83 MHz). Fig. \ref{F8}(a) depicts recovery curves at 67.7 MHz (under a magnetic field $H=15.90$ kOe) for selected temperatures.

 A departure from the single exponential behavior begins to appear below $T \sim 2.5$ K, and the recoveries of longitudinal magnetization display a double exponential variation of the form 
 \begin{equation} \label{eq5}
 	1-\frac{M(t)}{M(0)} = [Ae^{-(t/T_{1f})}+(1-A)e^{-(t/T_{1s})}],
 \end{equation}
 where, $1/T_{1f}$ corresponds to the faster rate with a fraction  $A$ and $1/T_{1s}$ is the slower component. Here, the fast and slow components in spin-lattice relaxation rate probably arise from protons at two different Wyckoff sites which might have different hyperfine couplings. In the temperature range $2.5\leq T\leq 1.5$ K, the value of $A$ is about 0.50, independent of temperature.

 Fig. \ref{F8}(b) shows the $T$-dependence of $1/T_1$ at 67.7 MHz (15.90 kOe) (for $0.09-125$ K), 101.5 MHz (23.85 kOe) (for $1.8-110$ K) and 194.83 MHz (45.79 kOe) (for $3-85$ K). Above 10 K, $1/T_1$ is roughly proportional to $T$, similar to the case of  $^1$H NMR in $k$-(BEDT-TTF)$_2$Hg(SCN)$_2$Cl \cite{pustogow2020impurity}. $1/T_1$ shows a remarkable decrease after going through  a hump at $T \sim$ 2.5 K. This appears to be due to the opening of a gap in the excitation spectrum, which can be due to the onset of a dimer singlet state.
 Using the $1/T_{1f}$ values at 67.7 MHz in the narrow temperature range 0.09 K – 2.4 K, the spin excitation gaps were evaluated to be $\Delta _1/k_{B}=5.97 (1.2) $ K and $\Delta _2/k_{B}=0.05 (0.03) $ K through the thermal-activation behavior $1/T_1 = A_1 \exp(-\Delta_1 /k_{B}T)+A_2 \exp(-\Delta_2 /k_{B}T)$ as clearly seen by solid line in Fig. \ref{F8}(b), where $A_1$ and $A_2$ are proportionality constants. The gap seen in $1/T_1$ in a field of 15.90 kOe might already be suppressed from its low-field value. We also attempted to fit the data using a single-exponential behavior with $\Delta/k_{B}=5.02 (0.9) $ K and a constant term (dashed line in Fig. \ref{F8}(b)). The constant term may indicate the presence of an additional relaxation mechanism at low temperature due to chain-end spins/residual impurities. The fitting of the $1/T_1$ data for the slower component yields similar values.
In quasi 1D chains, if the spin excitation spectrum has a structure such as more than one minimum, it will result in multiple peaks (hence multiple gaps) in the density of states of the excitation spectrum. This could arise if a low-energy two-magnon bound state is present \cite{private1}. Defects in 1D chains have also been known to give rise to a pseudogap in the spin excitations \cite{simutis2013}. A spin-Peierls transition would also produce gapped excitations. Further experiments which probe the magnetic excitations are required to understand the origin of the gapped state.

  The gap obtained here is different from the one obtained from zero-field $C_{mag}$  data in the range 0.07 K $\leq T$ $\leq$ 2.5 K. This could be due to the fact that the $q$-dependent excitations contribute to the two properties in different ways.
   The two-component fits to the recovery curves of the longitudinal nuclear magnetization yield that the coefficient $A$ deviates from 0.50 around $T=1$ K to about 0.30 in the temperature range 0.09 K $\leq T \leq 1$ K. 
   As seen in Fig. \ref{F8}(b), the faster component of $1/T_1$ is nearly 7.5 times greater than the slower one for 67.7 MHz at 0.09 K. 
 
\begin{figure}[ht]
	\begin{center}
		\includegraphics[width=0.95\columnwidth]{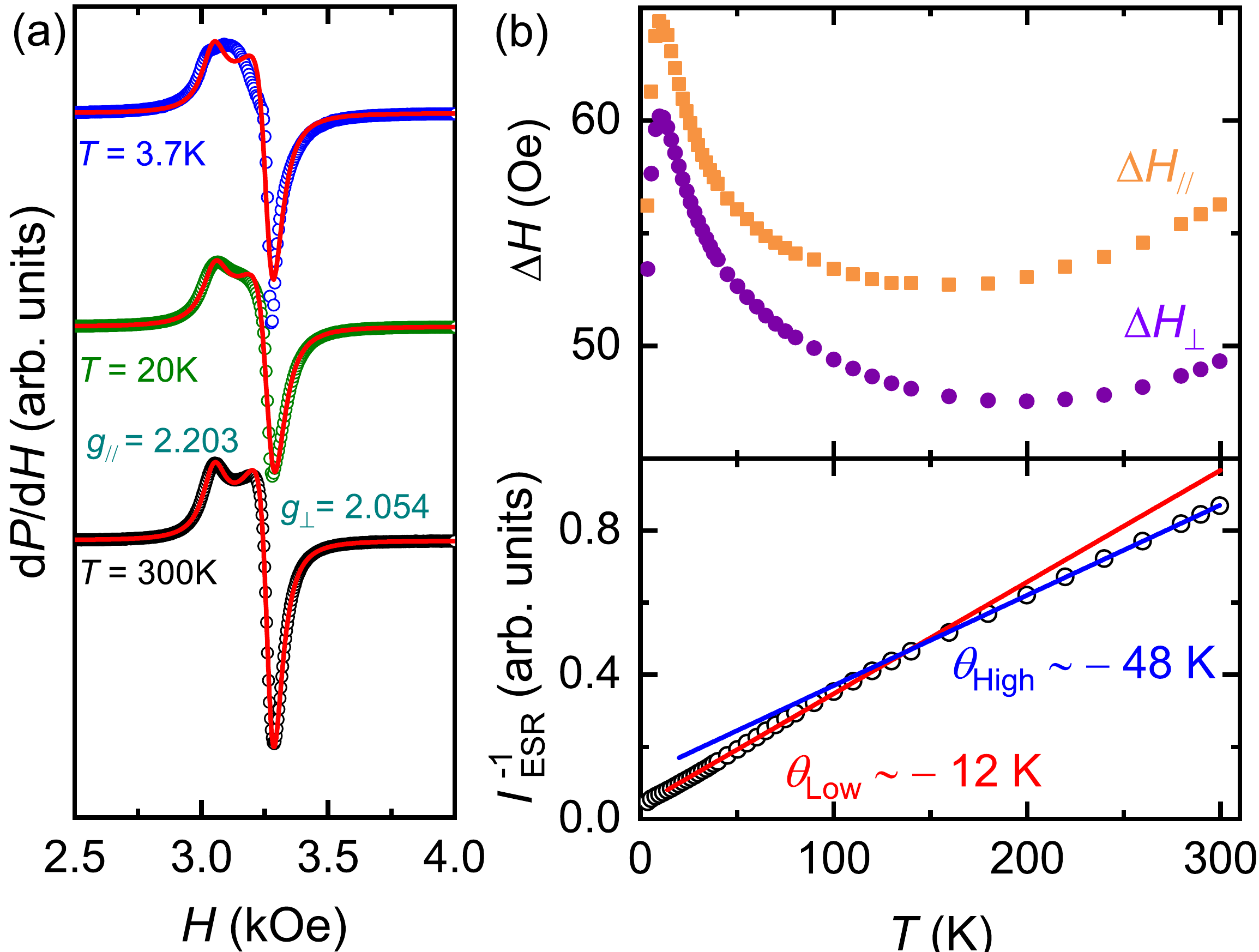}
		\caption{(a) X-band ESR spectra (symbols) at representative temperatures with fit to uniaxial powder-averaged Lorentzians (solid red lines). Fitted values of anisotropic $g$-values are indicative for the spectrum at $T=$ 300 K. (b) Temperature dependence of fitted parameters linewidth (top-frame) and inverse integrated ESR intensity with Curie-Weiss behavior as indicated by solid lines (bottom-frame).}
		\label{F9}
	\end{center}
\end{figure}
\subsection{Electron Spin Resonance}

Electron Spin Resonance is employed as a local probe technique to investigate the magnetic interactions between intrinsic Cu$^{2+}$  moments in Cu(Ampy)ClBr. The evolution of ESR spectra with temperature is shown in Fig. \ref{F9}(a). The spectra were fitted by a powder-averaged, uniaxial Lorentzian lineshape yielding in-plane and out-of-plane components of linewidth $\Delta H$ and $g$-factor. The ESR spectra are thus characterized by $g_\|$, $g_\bot$ and $\Delta H_{\|}$, $\Delta H_{\bot}$ along the \textit{c}-axis and in the \textit{ab}-plane crystallographic directions, respectively. 
At $T=300$~K, we find $g_\|=2.203$ and $g_\bot=2.054$, corresponding to an average value $g_{\rm avg} = \sqrt{(g_\|^2+2g_\bot^2)/3} = 2.105$. 
As can be seen in Fig. \ref{F9}(b), top-frame, the temperature dependencies of the linewidths show a continuous broadening below $T\simeq160$~K, indicating the enhancement of Cu$^{2+}$ spin correlations \cite{sana2024}.
A remarkable drop of the linewidth appears below around $T = 10$~K. At the same time, the fit quality is strongly reduced leading to narrow components in the lineshape (see the spectrum at $T=3.7$~K for instance). While the origin of this lineshape development remains unclear, the effect of short-range magnetic correlations could be relevant as well \cite{cinthia2022}. Above $T \simeq160$~K both components of $\Delta H$ increase. Theoretical studies suggest that a phonon modulation of the anisotropies at temperatures comparable with the Debye temperature ($\theta_{\rm D} \approx 100$~K) leads to a linear increase in $\Delta H$ \cite{soos1977, willett1981}. The integrated ESR intensity $I_{\rm ESR}$ follows a Curie-Weiss behavior with a Weiss temperature $\theta_{\mathrm{Low}}=-12(1)$~ K in the temperature range $12 \le T \le 150$~K as shown in Fig. \ref{F9}(b), bottom-frame. This value of $\theta_{CW}$ is consistent with the one obtained from magnetic susceptibility. Also, $\theta_{CW}$ changes to $\theta_{\mathrm{High}}=-48(2)$ K if the fit is carried out in the temperature range $160 \le T \le 300$~K. This effect may originate from uncertainties in lineshape fitting. In this respect one should be aware of the structural disorder caused by carbon C(6) atom in Ampy rings over two positions related by a mirror plane, which do not influence the structure of magnetic zigzag chains \cite{tarasenko2023, cinthia2022}.

\subsection{Muon spin relaxation}

\begin{figure*}
	\begin{center}
		\includegraphics[width=2.05\columnwidth]{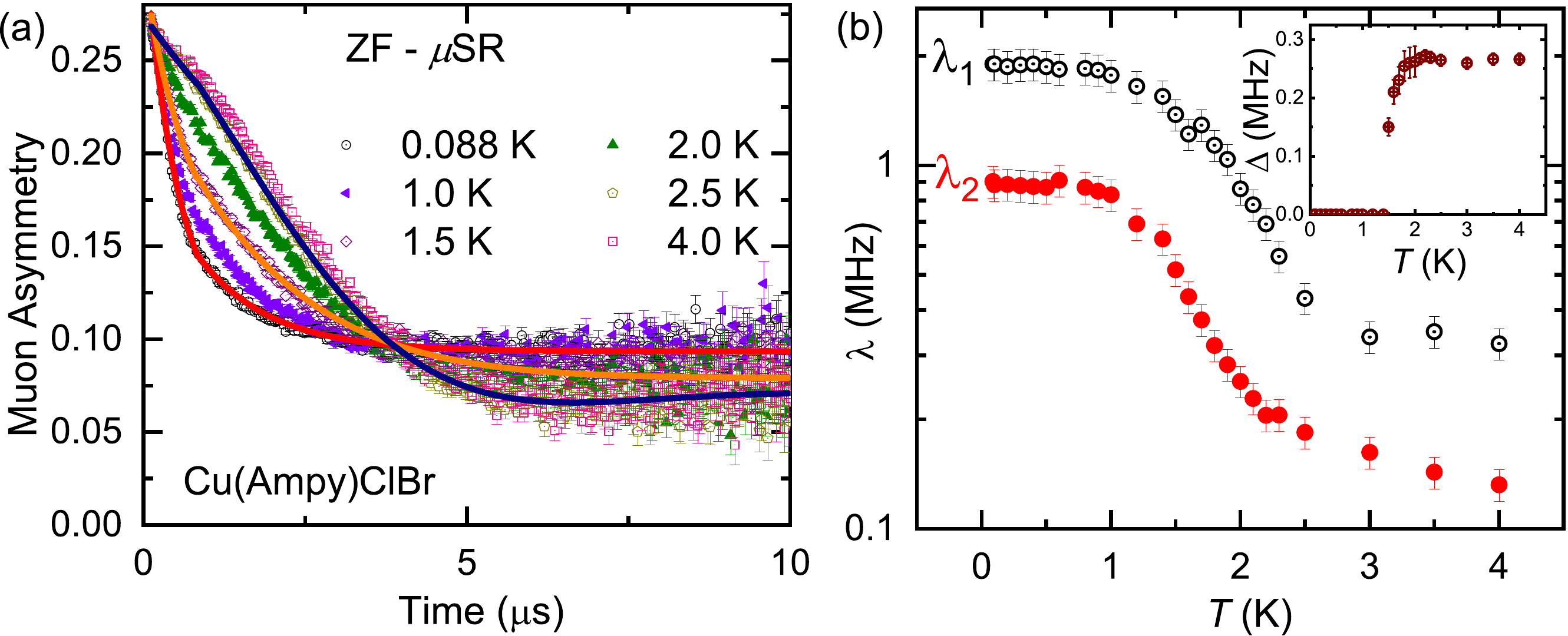}
		\caption{(a) The time evolution of the ZF-$\mu$SR asymmetry spectra at selected temperatures. Solid lines are fits to the data as described in the text. (b) The temperature dependence of the muon spin relaxation rates, $\lambda_{1}$ and $\lambda_{2}$ are extracted from fits to the data. Inset: Temperature dependence of the parameter $\Delta$.}
		\label{F10}
	\end{center}
\end{figure*}

To investigate the spin dynamics in Cu(Ampy)ClBr, we conducted ZF and LF $\mu$SR measurements for a powder sample down to 0.088 K. Fig. \ref{F10}(a) shows the ZF-$\mu$SR spectra at selected temperatures. The muon depolarization as a function of time is  dominated by a Gaussian-like relaxation in the temperature range from 4 K to 2.5 K, featuring a dip around $\sim 6$ $\mu$s followed by a recovery of the asymmetry to about 1/3$^{rd}$ the original value at  longer times \cite{blundell2021}. Such behavior is typically observed in systems with randomly oriented static internal fields, where the field distribution is Gaussian due to nuclear moments.
Upon lowering the temperature further, the shape of muon depolarization curves changes to an  exponential-like behavior. Below $T\sim 1.5$ K, the dip is completely absent. 
The Gaussian damping is typical of quasi-static nuclear moments. In addition, the  fluctuations of electronic moments lead to an exponential relaxation of the muon spin polarization. 

The muon depolarization  exhibits neither coherent  oscillations nor a loss of initial asymmetry down to 0.088 K, ruling out the transition to a magnetic long range ordering (LRO) state. Any static random field distribution was also excluded due to the lack of long time recovery to one-third of the initial asymmetry, indicating that the relaxation was caused by dynamic effects.

We find that the muon asymmetry below about 1.5 K cannot be fit to a single exponential. We then extract the faster component of the muon depolarization rate $\lambda_1$ from the initial exponential variation. In short, the following function is used in the small-time  region of the asymmetry variation,  
\begin{equation} \label{eq6}
	A(t) = A_{1} exp(-\lambda_{1}t)  + A_{0}
\end{equation} 
where $A_0$ is the background term and $A_1$ is the weight of the relaxing term. 
Likewise, the slower rate ($\lambda_2
$) is obtained from the variation in the long-time ($t > 1$ $\mu$s) part of the asymmetry curve to an equation similar to the one above.  For $T > 1.5$ K, we need to multiply the exponential in the above equation with a Kubo-Toyabe function  to fit the data.  $G_{ZF}^{KT}$ is the ZF  Kubo-Toyabe function reflecting the Gaussian distribution of randomly oriented or quasistatic local magnetic fields at the muon sites given by,
\begin{equation} \label{eq7}
	G_{ZF}^{KT} = \frac{2}{3} (1-\Delta^2t^2) exp(-\frac{1}{2}\Delta^2t^2) + \frac{1}{3}
\end{equation}
\\
where $\Delta / \gamma_\mu$ describes the root-mean-square (rms) width of the Gaussian distribution and $\gamma_\mu = 2\pi \times 135.53$ MHz/T is the muon gyromagnetic ratio.

The parameter $\Delta$ is the same for both fitting procedures and found to be temperature independent down to 2 K, with $\Delta \approx 0.26$ MHz corresponding to an rms quasistatic local field $\Delta H\approx3$ Oe, which is a typical value of the nuclear dipolar field distribution.

As the temperature is lowered, the value of $\Delta$ eventually drops to zero below $T \sim$ 1.5 K (see inset of Fig. \ref{F10}(b)).
\begin{figure*}
	\begin{center}
		\includegraphics[width=2.05\columnwidth]{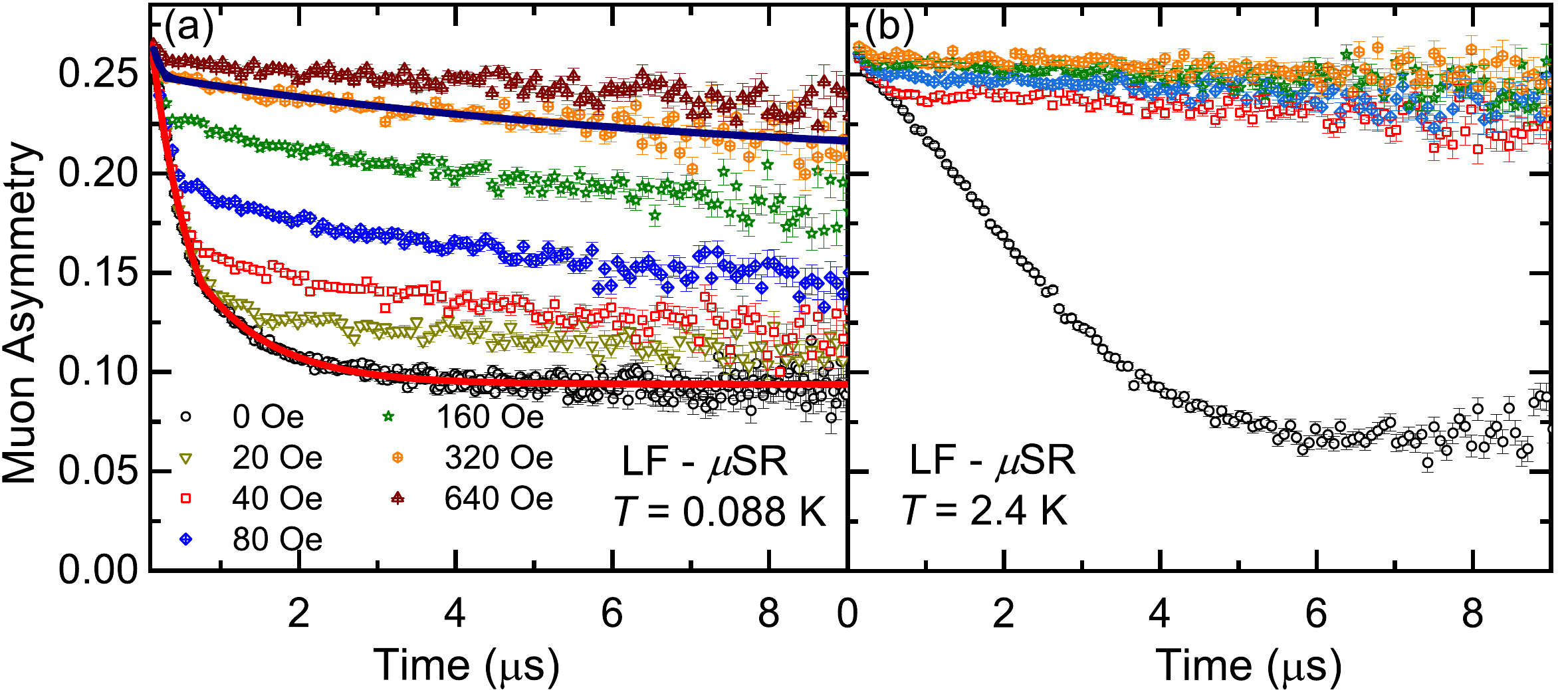}
		\caption{Muon asymmetry as a function of decay time at (a) $T=$ 0.088 K and (b) 2.4 K for various longitudinal fields. Solid lines represents fittings described in the text.}
		\label{F11}
	\end{center}
\end{figure*}

This basically means that relaxation from local electron moment fluctuations totally dominate over the relaxation from nuclear moments.
Fig. \ref{F10}(b) shows the temperature dependence of the muon spin-relaxation rates.  At temperatures above 2 K, both $\lambda_{1} (T)$ and $\lambda_{2} (T)$ display no temperature dependence. With decreasing temperature below 2 K, they increase gradually and then become constant below 1 K and down to 0.088 K. This is evidence that the observed relaxations arise from dynamical spin fluctuations, which persist down to 0.088 K. The temperature dependence of the relaxation rate is similar to that in quantum spin liquid systems \cite{balz2016physical}.

We have measured the LF-$\mu$SR response of Cu(Ampy)ClBr in different fields. In an applied LF, the nuclear contribution to the muon spin relaxation is expected to be fully quenched for fields higher than about  5$\Delta H$,
while the required LF is much higher in case of dynamically fluctuating spins. Fig. \ref{F11}(a) presents the muon spin polarization at the base temperature of 0.088 K in different LFs.  Remarkably, substantial relaxation is clearly observed in an LF of 160 Oe ($\sim 50 \Delta H$) and 
the asymmetry is fully recovered above 320 Oe ($> 100 \Delta H$), which corroborates the existence of the electronic nature of fluctuations without static ordering down to 0.088 K. On the other hand, the significant change in muon asymmetry between zero field and applied longitudinal fields at 2.4 K shows decoupling of muons at LF of 40 Oe ($> 10 \Delta H$) 
(see Fig. \ref{F11}(b)). A similar muon response has been observed in $S$ = 1/2 triangular quantum magnet, Cu($1,3-$diaminopropane)Cl$_2$ \cite{tarasenko2023}.
\begin{figure}[ht]
	\begin{center}
		\includegraphics[width=0.95\columnwidth]{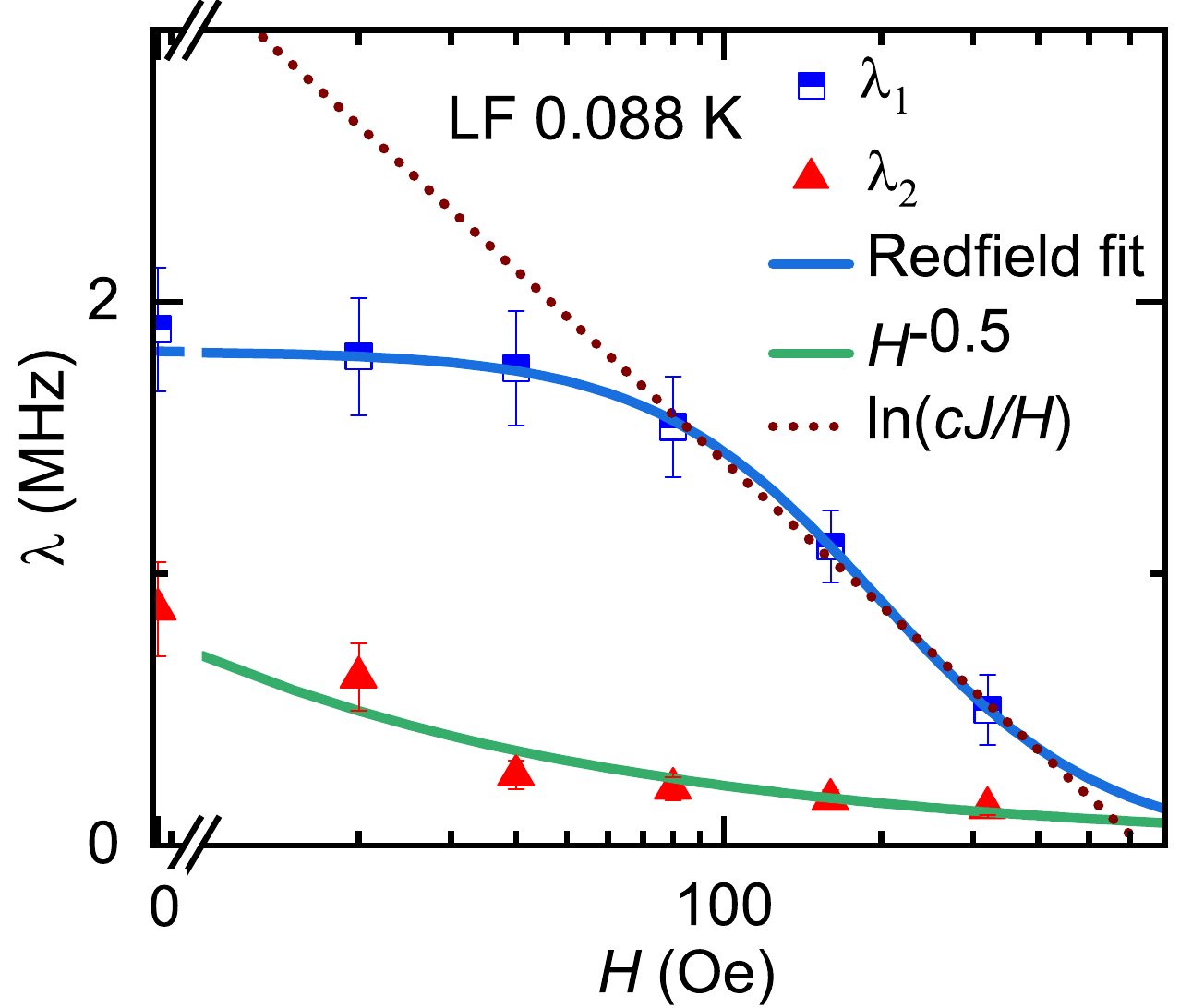}
		\caption{Longitudinal-field dependence of the muon spin relaxation rates $\lambda_{1}$ and $\lambda_{2}$  measured at $T=0.088$ K. Also shown are expected variations in case of ballistic ($\propto$ ln$(cJ/H)$) and diffusive ($\propto H^{-0.5}$) transport of spinons as also the Redfield variation.}
		\label{F12}
	\end{center}
\end{figure}

The LF data are well described by an exponential variation, as given in Eq. \ref{eq6}, for both short-time region and the long-time ($t > 1$ $\mu$s) part of the asymmetry curve (see Fig. \ref{F11}(a)) respectively. To extract the faster and slower component of the muon depolarization rate, we have used the same techniques as described in case of ZF-$\mu$SR for the base temperature.
 In Fig. \ref{F12}, we exhibit the field dependence of the relaxation rates $\lambda_{1}$ and $\lambda_{2}$. The longitudinal-field dependence of $\lambda$  gives the opportunity to understand the nature of spin excitations (whether ballistic or diffusive) since the spin correlation functions are governed by their transport behavior.

The spin autocorrelation functions have different spectral densities where  $\mathcal{S} (\omega) $ varies with frequency $\omega$ as $\mathcal{S} (\omega) \sim \omega ^{-1/2}$ for 1D diffusive transport and $\mathcal{S} (\omega) \sim$ ln$(J/ \omega)$ for ballistic spin transport \cite{pratt1997,pratt2006}. Consequently, $H$ dependence of $\lambda$ obeys the $H^{-1/2}$ power-law  for 1D diffusion and ln$(J/H)$  for ballistic spin transport \cite{blundell2021}. As illustrated in Fig. \ref{F12}, the slower muon spin relaxation rate $\lambda_2(H)$ at $T = $ 0.088 K is well reproduced by the 1D spin diffusion model. Such power-law decaying spin correlations are observed in various materials with spin-1/2 HAF chain, pyrochlore, kagome, and triangular lattices \cite{pratt2006, blundell2021, lee2020, kermarrec2014,leeW2024, keren2004}. However, as shown in the Fig. \ref{F12},  ballistic motion provides a poor description of the $\lambda(H)$ data. We suggest that muons sites near the  chain-center sense a diffusive motion of spinons as the presence of Cl/Br randomness in Cu(Ampy)ClBr might prevent a ballistic transport of spinons.
On the other hand, the faster rate $\lambda _1$ is well described by the Redfield relation \cite{Redfield1957, hayano1979}
\begin{equation} \label{eq8}
	\lambda _1 (H) = \frac{{2\gamma_{\mu}^{2} \langle B_{loc}^{2} \rangle \nu}}{(\gamma_{\mu}H)^2+\nu ^2}
\end{equation}
where, $\nu$ is the fluctuation frequency (related to the correlation time $\tau= 1/\nu$) and $B_{loc}$ is the time average of the fluctuating amplitude of local internal field. We infer $\nu \approx$ 16.8 MHz and $B_{loc} \approx$ 46 Oe.  We suggest that this contribution arises from muon sites near the chain-ends where the relaxation is controlled by the fluctuations of the staggered moments released near chain-ends.

\section {Discussion}

We now consider all the results obtained by us and examine their implications to the big picture. We set out to introduce disorder in the exchange coupling through Cl mixing at the Br site in the Cu(Ampy)Br$_2$ chain system with the intention of suppressing order and possibly driving it towards some exotic ground state.   In the phase diagram given in Ref. \cite{uematsu2021frustration}, an interplay of the randomness parameter and a frustrating next-nearest neighbor coupling can give rise to either an unfrustrated random-singlet, a frustrated random-singlet, or a dimer-singlet ground state. No indication of any long-range magnetic ordering was found in any of our measurements indicating the good one-dimensional nature of our system. While there is a hint of a broad maximum  in the susceptibility data (expected for one-dimensional systems) between 1 K and 10 K (Fig. \ref{F3}(a)), a Curie-like upturn appears to dominate at low temperature which overshadows any possible intrinsic decrease. We have attempted to fit the susceptibility and the zero-field magnetic specific heat data using the Johnston expression for the spin-1/2 HAF chain, but it clearly fails to fit the data below broad maximum, suggesting deviations from the uniform HAF model. 
Since we do not have a microscopic $J_1-J_2$ model Hamiltonian to describe the chain susceptibility and specific heat, it is challenging to fit the entire data set while accounting for the gapped behavior observed at low temperatures.
The magnetic specific heat data show a sharp decrease below  $T \sim$ 2.5 K, and the data are consistent with gapped excitations. The hump around $T \sim$ 2.5 K in $C_{mag}$ itself might indicate a deviation from the uniform chain behavior which remains even in high field. The presence of  gap signature is also evident in the temperature dependence of  $^1$H NMR  $1/T_1$ below about 2.5 K (Fig. \ref{F8}(b)).
It is worth noting that the spin gap value from NMR spin-lattice relaxation rate $1/T_1$ is almost twice as large as the gap value derived from the specific heat. A similar observation was reported in case of the 1D chain compound, SrNbO$_{3.41}$, where the gap from NMR 1/T$_1$ was twice the size of the gap obtained from transport properties, i.e., resistivity \cite{kuntscher2002, weber2001}.
The gap determined from NMR relaxation rate ($1/T_1$) (which probes $q$-integrated dynamical susceptibility) can be different from that obtained from, say,- uniform susceptibility $\chi$($q=0$) or specific heat.

Our $\mu$SR results throw more light on various aspects of this one-dimensional system. While the absence of static magnetism down to 0.088 K (much lower than the nominal exchange coupling of 18 K) is clear, our zero-field and longitudinal field $\mu$SR measurements support a dynamical ground state in Cu(Ampy)ClBr. There is a  levelling-off of the $T$-dependence of $\lambda$ below $T\sim$ 1 K. 
The magnetic field dependence of the muon relaxation rate is expected to probe the motion of spinons in  one-dimensional systems; distinct variations are predicted for ballistic and diffusive motion. Indeed, such behavior has been reported in Refs. \cite{pratt2006, blundell2021}. 
A gapped ground state can still give rise to confined spinons and in the timescale of $\mu$SR, we can still observe their motion.
Given that we have exchange interaction disorder due to Br/Cl mixing, it is perhaps not surprising that we do not see a variation expected for ballistic transport and rather a diffusive description fits the data. A second (faster) component in the muon relaxation rate is seen which follows the Redfield behavior. We attribute this faster component to muon sites near chain-ends where the relaxation is largely driven by chain-end moment fluctuations. 

There are various scenarios that can lead to the opening of a gap in the spin excitation spectrum. One possibility, which we have pointed out, is associated with a suitable combination of the randomization parameter and the ratio of the next-nearest-neighbour to the nearest-neighbour coupling ($J_2/J_1$) \cite{uematsu2021frustration}. A second possibility is the opening of a pseudogap due to defects in the $S=1/2$ chain. The value of the pseudogap, as determined from inelastic neutron scattering, would then depend on the defect concentration \cite{simutis2013}. This could also give rise to spinon confinement, as inferred by us from the $\mu$SR data (see Fig. \ref{F12}). Another possibility is a Spin-Peierls transition, which would be accompanied by a structural transition leading to dimerization \cite{hirota1994,pouget1994}. The first two possibilities do not require any structural transition, while the third one does. In fact, in Ni-doped SrCuO$_2$, it was concluded that the gap observed in inelastic neutron scattering was not due to a Spin-Peierls transition as no structural transition was detected \cite{simutis2013}. In our case, structural measurements down to low temperatures would be needed to definitively rule-out the Spin-Peierls possibility.
 
\section {Conclusion}
In this work, we have successfully synthesized and studied the magnetic and thermodynamics properties of $S=1/2$ compound Cu(Ampy)ClBr, employing various bulk and local probe techniques down to 0.06 K. The Cu$^{2+}$ ions constitute a zigzag chain structure with moderate antiferromagnetic (AFM) interaction between the Cu-local moments. The average AFM exchange interaction is $J/k_{B} \sim 18(2)$ K from the broad maxima in the susceptibility. Cu(Ampy)ClBr does not exhibit any static magnetic ordering, instead shows the presence of persistent dynamics of Cu$^{2+}$ moments down to 0.06 K, which is well below the CW temperature $\theta_{CW}$ of $-9$ K. 
 
Our data reveal the presence of a  gapped ground state. This could be due to a dimer-singlet state arising from an interplay of the randomness parameter and  frustrating next-nearest neighbor interaction or pseudogap resulting from defects in the spin-chain. A diffusive motion of spinons is detected through the field variation of the $\mu$SR relaxation rate in addition to a contribution with a Redfield variation which is suggested to originate from muon sites near the chain-ends. This work should motivate others to prepare single crystals and examine the excitations in this system through, for example, inelastic neutron scattering in deuterated samples to try to arrive at a microscopic model Hamiltonian. Thermal transport is another suggested direction to probe spinon excitations. \\

\section*{Acknowledgments}
We thank DST-SERB for funding under the DST-CRG scheme (CRG/2021/003024). We acknowledge the use of various Central Facilities at IIT Bombay. We thank ISIS Facility for beam time allocation [RB2368023]. S.N. and A.V.M. would like to thank the Department of Science and Technology (DST), India, for access to the experimental facility and financial support for the experiment conducted at ISIS muon source and Jawaharlal Nehru Centre for Advanced Scientific Research (JNCASR) for managing the project.  S.N. would like to acknowledge the funding support for Chanakya Postdoctoral fellowship (Registration No. CPDF/2021-22/01) from the National Mission on Interdisciplinary Cyber Physical Systems, of the DST, Govt. of India through the I-HUB Quantum Technology Foundation. A.S. is thankful to the DST for Grant No. DST/NM/TUE/QM-10/2019 (G)/3. R.K. is thankful to the Council of Scientific and Industrial Research (CSIR) for Ph.D. fellowship (Grant No. 09/733(0263/2019-EMR-I)). This work was partially supported by the Deutsche Forschungsgemeinschaft (DFG) within the Transregional Collaborative Research Center TRR 360 ``Constrained Quantum Matter", Project No. 492547816 (Augsburg, Munich, Stuttgart, Leipzig). Work at Duke University has been supported by the NSF grant DMR2218058. We would like to thank M. Baenitz and G. Baskaran for enriching discussions.

%
%
%
%
%
%
%
%
%
%
%
%
%
%
%
%
%
%
%
%

\bibliography{ref}

\end{document}


\title{Supplemental Material: Observation of a gapped phase in the one-dimensional $S\!=\!\frac{1}{2}$ Heisenberg antiferromagnetic chain Cu(Ampy)ClBr}
\author{Saikat Nandi}
\email{saikatnandi9@gmail.com}
\affiliation{Department of Physics, Indian Institute of Technology Bombay, Mumbai 400076, India}
\author{Monika Jawale}
\affiliation{Department of Physics, Indian Institute of Technology Bombay, Mumbai 400076, India}
\author{Sanjay Bachhar}
\affiliation{Department of Physics, Indian Institute of Technology Bombay, Mumbai 400076, India}
\affiliation{Department of Physics, Duke University, Durham, North Carolina 27708, USA}
\author{Rahul Kumar}
\affiliation{School of Advanced Materials, and Chemistry and Physics of Materials Unit, Jawaharlal Nehru Centre for Advanced Scientific Research, Bangalore-560064, India}
\author{Marlis Schuller}
\affiliation{Experimental Physics V, Center for Electronic Correlations and Magnetism, University of Augsburg, D-86135 Augsburg, Germany}
\author{Rabindranath Bag}
\affiliation{Department of Physics, Duke University, Durham, North Carolina 27708, USA}
\author{J. M. Wilkinson}
\affiliation{ISIS Pulsed Neutron and Muon Source, STFC Rutherford Appleton Lab, Harwell Campus, Didcot OX11 0QX, Oxon, England}
\author{Jörg Sichelschmidt}
\affiliation{Max Planck Institute for Chemical Physics of Solids, 01187 Dresden,
Germany}
\author{A. Sundaresan}
\affiliation{School of Advanced Materials, and Chemistry and Physics of Materials Unit, Jawaharlal Nehru Centre for Advanced Scientific Research, Bangalore-560064, India}
\author{Sara Haravifard}
\affiliation{Department of Physics, Duke University, Durham, North Carolina 27708, USA}
\author{N. Büttgen}
\affiliation{Experimental Physics V, Center for Electronic Correlations and Magnetism, University of Augsburg, D-86135 Augsburg, Germany}
\author{A.V. Mahajan} 
\email{mahajan@phy.iitb.ac.in}
\affiliation{Department of Physics, Indian Institute of Technology Bombay, Mumbai 400076, India}


\maketitle

Herein we report details of various measurements (XRD, magnetization, specific heat) performed by us on Cu(Ampy)X$_2$ (X= Cl, Br). 

\subsection{XRD}

\renewcommand{\thefigure}{S\arabic{figure}}  
\setcounter{figure}{0}                       
\begin{figure}[ht]
	\begin{center}
		\includegraphics[width=0.75\columnwidth]{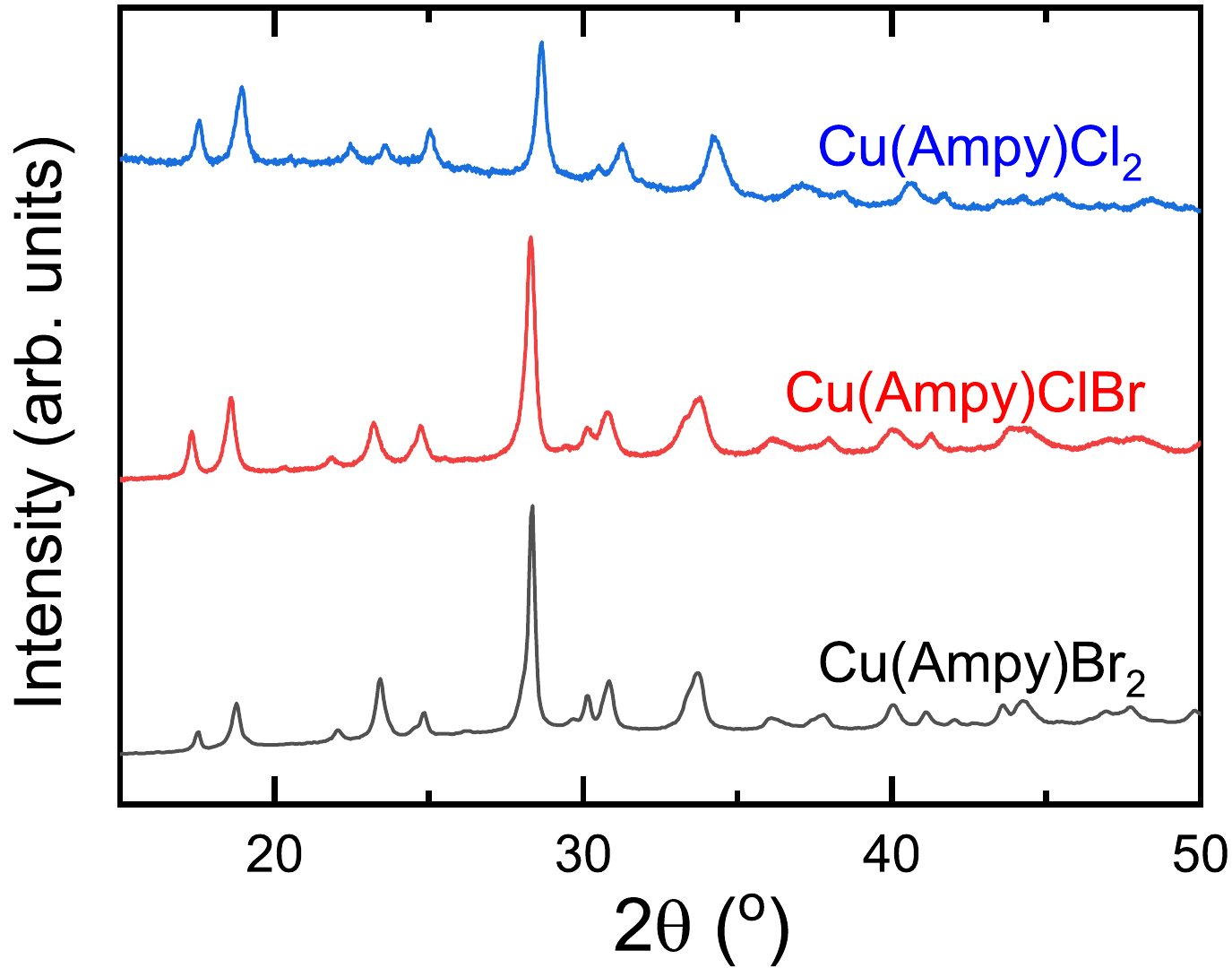}
		\caption{Powder x-ray diffraction pattern for  Cu(Ampy)Cl$_2$, Cu(Ampy)ClBr, and Cu(Ampy)Br$_2$ at 300 K.}
		\label{fig:S1}
	\end{center}
\end{figure}
   
  Polycrystalline samples of Cu(Ampy)Cl$_2$ and Cu(Ampy)Br$_2$ were synthesized using the same technique employed for Cu(Ampy)ClBr. The phase purity of these compounds were checked by doing powder x-ray diffraction (XRD) measurements at room temperature with a high resolution Rigaku diffractometer using Cu-K$_\alpha$ radiation ($\lambda$ = 1.5406 Å) (see Fig. \ref{fig:S1}). Cu(Ampy)Cl$_2$, Cu(Ampy)ClBr, and Cu(Ampy)Br$_2$ crystallize in the monoclinic space group $P$2$_1/$$m$.
  
\subsection{Magnetization}

\begin{figure}[ht]
	\begin{center}
		\includegraphics[width=0.75\columnwidth]{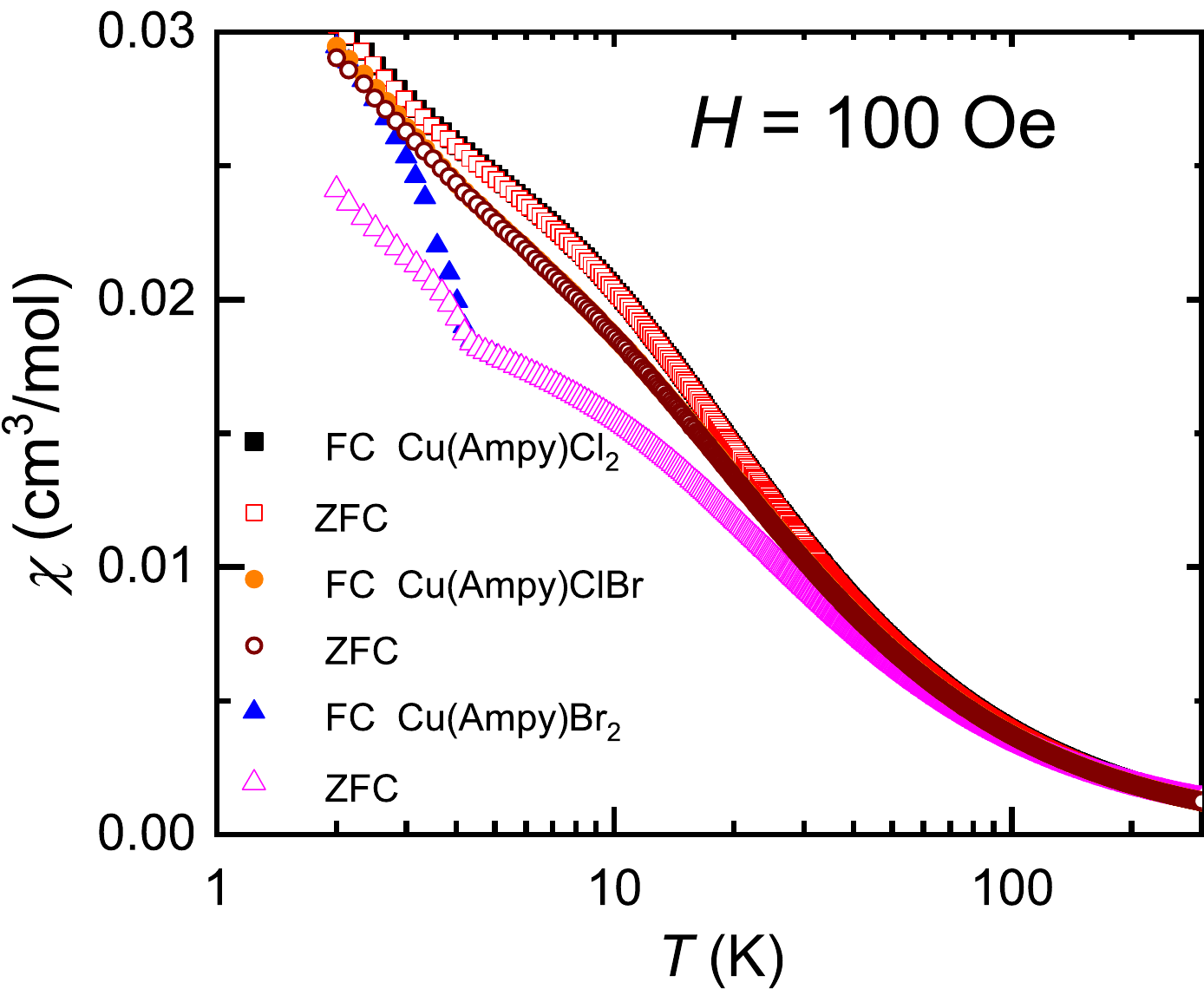}
		\caption{(a) Temperature dependence of the static magnetic susceptibility $\chi(T)$, for Cu(Ampy)Cl$_2$, Cu(Ampy)ClBr, and Cu(Ampy)Br$_2$, measured at $H=$ 100 Oe under  field cooled (FC) (close symbols) and zero field cooled (ZFC) (open symbols) conditions, down to 2 K on a semi-log scale. A bifurcation between the ZFC and FC $\chi(T)$ curves in Cu(Ampy)Br$_2$ is observed around 4 K.}
		\label{fig:S2}
	\end{center}
\end{figure}

The temperature dependence of the zero-field-cooled (ZFC) and field-cooled (FC) magnetic susceptibility, $\chi (T)$, of Cu(Ampy)Cl$_2$, Cu(Ampy)ClBr, and Cu(Ampy)Br$_2$, measured down to 2 K in an applied field of $H =$ 100 Oe, is shown in Fig. \ref{fig:S2}. At high temperatures, $\chi (T)$ follows the standard paramagnetic behavior. With decreasing temperature, it exhibits a broad and weak maximum around $T \simeq$ 9 K, qualitatively similar to other chain systems, characteristic of low-dimensional systems with short-range ordering. The $\chi_{max}$ values for Cu(Ampy)Cl$_2$ and Cu(Ampy)Br$_2$ at around 9 K are somewhat higher than those previously reported in Ref. \cite{OConnor1985, kikuchi2000}, probably 
due to the presence of end-chain spins, perhaps due to the smaller crystallite sizes in our samples. The bulk dc susceptibility of Cu(Ampy)Br$_2$ shows a ZFC/FC bifurcation around 4 K, clearly indicating a transition to long-range magnetic order. 
	
		
		\subsection{Specific Heat}
		
\begin{figure}[ht]
		\begin{center}
			\includegraphics[width=0.75\columnwidth]{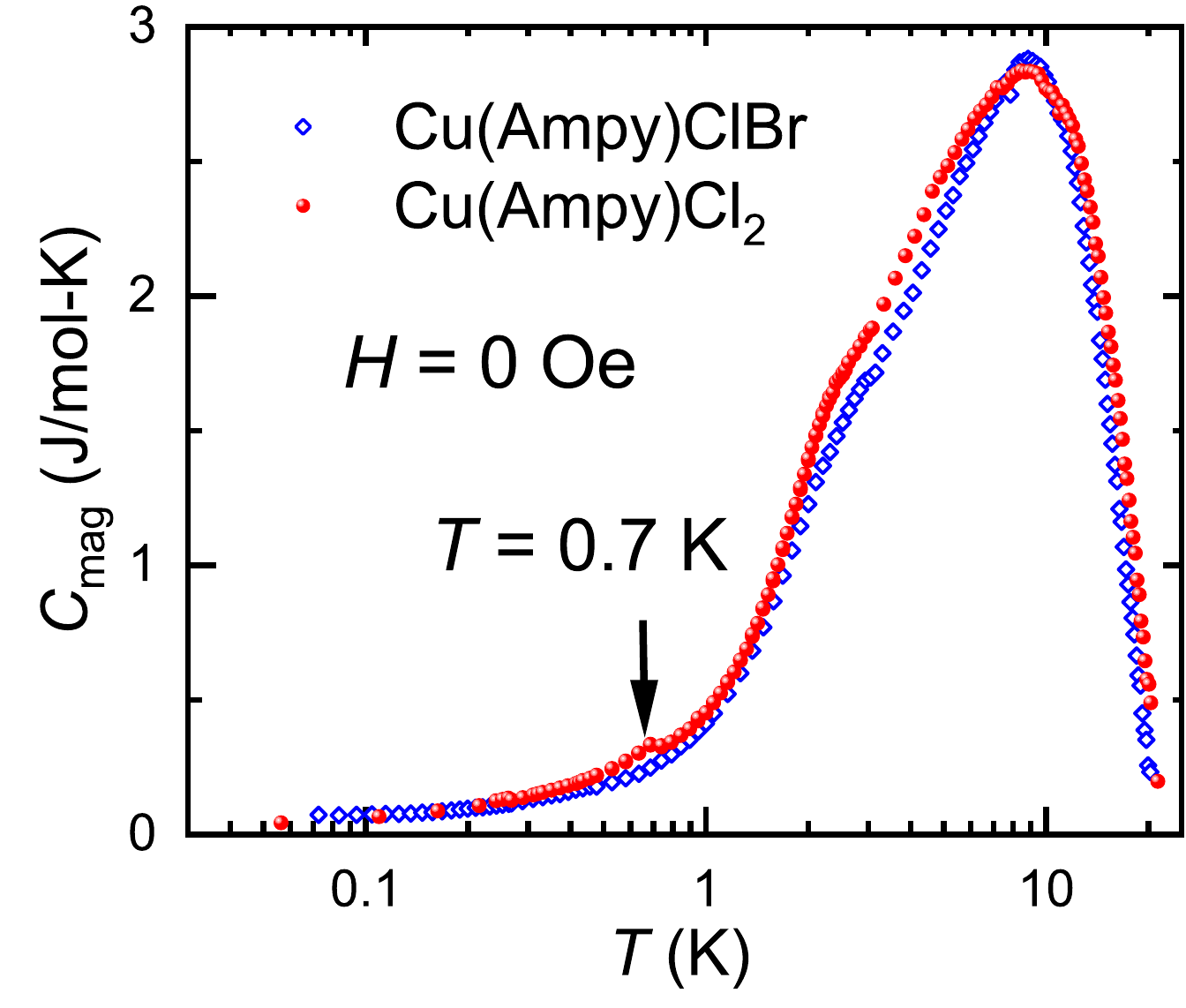}
			\caption{(b) Magnetic contribution to the specific heat ($C_{mag}$) for Cu(Ampy)ClBr and Cu(Ampy)Cl$_2$ as a function of temperature in zero field. }
			\label{fig:S3}
		\end{center}
	\end{figure}

Fig. \ref{fig:S3} shows the temperature dependence of magnetic specific heat, $C_{mag}$, for Cu(Ampy)Cl$_2$, Cu(Ampy)ClBr in zero magnetic field. The $C_{mag}$ \textit{vs} $T$ exhibits a broad hump at around $\sim 9$ K, which is typical for low-dimensional systems. Additionally, the specific heat data for Cu(Ampy)Cl$_2$ display a sharp anomaly around 0.7 K.

%
%
%
%
%
%
%
%
%
%
%
%
%
%
%
%
%
%
%
%

\bibliography{ref}